\documentclass[journal]{IEEEtran}
\usepackage{amsmath}
\usepackage{amssymb}
\usepackage{amsthm}
\usepackage{amsfonts}
\usepackage{epsfig}
\usepackage{psfrag}
\usepackage{graphicx}
\usepackage{mathrsfs}
\usepackage{color}
\usepackage{cite}
\usepackage{textcomp} 

\usepackage[colorlinks=true,linkcolor=black,anchorcolor=black,citecolor=black,filecolor=black,menucolor=black,runcolor=black,urlcolor=black]{hyperref}
\usepackage{multirow}

\usepackage{physics} 
\usepackage{empheq}
\usepackage{authblk}
\usepackage[normalem]{ulem} 
\usepackage{mathtools} 

\theoremstyle{definition}
\newtheorem{assumption}{Assumption}
\newtheorem{proposition}{Proposition}
\newtheorem{theorem}{Theorem}
\newtheorem{lemma}{Lemma}
\newtheorem{remark}{Remark}

\usepackage{algorithm}
\usepackage{algorithmicx}
\usepackage{algpseudocode}

\ifCLASSINFOpdf
\else
\fi
\usepackage{empheq}

\begin{document}
%
\title{Training Time Minimization for Federated Edge Learning with Optimized Gradient Quantization and Bandwidth Allocation}

\author{Peixi~Liu,~Jiamo~Jiang,~Guangxu~Zhu,~Lei~Cheng,~Wei~Jiang,~Wu~Luo,~Ying~Du,~Zhiqin~Wang
\thanks{Peixi Liu is with State Key Laboratory of Advanced Optical Communication Systems and Networks, School of Electronics, Peking University, Beijing, China, and Shenzhen Research Institute of Big Data, Shenzhen, China (e-mail: liupeixi@pku.edu.cn).} \thanks{Jiamo Jiang, Ying Du and Zhiqin Wang are with China Academy of Information and Communications Technology (e-mail:  jiangjiamo, duying1, zhiqin.wang@caict.ac.cn).} \thanks{Guangxu Zhu is with Shenzhen Research Institute of Big Data, Shenzhen, China (e-mail: gxzhu@sribd.cn).} \thanks{Lei Cheng is with College of Information Science and Electronic Engineering, Zhejiang University, Hangzhou, China (e-mail:  lei\_cheng@zju.edu.cn)}\thanks{Wei Jiang and Wu Luo are with State Key Laboratory of Advanced Optical Communication Systems and Networks, School of Electronics, Peking University, Beijing, China (e-mail: jiangwei, luow@pku.edu.cn).}}

\maketitle

\begin{abstract}
Training a machine learning model with federated edge learning (FEEL) is typically time-consuming due to the constrained computation power of edge devices and limited wireless resources in edge networks.
In this paper, the training time minimization problem is investigated in a quantized FEEL system, where the heterogeneous edge devices send quantized gradients to the edge server via orthogonal channels.
In particular, a stochastic quantization scheme is adopted for compression of uploaded gradients, which can reduce the burden of per-round communication but may come at the cost of increasing number of communication rounds. 
The training time is modeled by taking into account the communication time, computation time and the number of communication rounds.
Based on the proposed training time model, the intrinsic trade-off between the number of communication rounds and per-round latency is characterized.
Specifically, we analyze the convergence behavior of the quantized FEEL in terms of the optimality gap. 
Further, a joint data-and-model-driven fitting method is proposed to obtain the exact optimality gap, based on which the closed-form expressions for the number of communication rounds and the total training time are obtained. 
Constrained by total bandwidth, the training time minimization problem is formulated as a joint quantization level and bandwidth allocation optimization problem.
To this end, an algorithm based on alternating optimization is proposed, which alternatively solves the subproblem of quantization optimization via successive convex approximation and the subproblem of bandwidth allocation via bisection search.
With different learning tasks and models, the validation of our analysis and the near-optimal performance of the proposed optimization algorithm are demonstrated by the experimental results.
\end{abstract}


%
\IEEEpeerreviewmaketitle

\section{Introduction} \label{sec:introduction}
The evolution of wireless networks from 1G to 5G-Advanced has witnessed a paradigm shift from connecting people targeting human-type communications towards connecting intelligence targeting machine-type communications to attain the vision of artificial intelligence of things (AIoT). This has driven the rapid development of an emerging area called \textit{edge intelligence}, sitting at the intersection of the two disciplines, namely artificial intelligence (AI) and wireless communications \cite{Letaief2019CM-6G,Park2019Proceed_edge}. 
In edge intelligence, AI technologies are pushed towards the network edge so that the edge servers can quickly access real-time data generated by edge devices for fast training and real-time inference \cite{Zhu2020CM-Edge}.
A promising framework for distributed edge learning, called \textit{federated edge learning} (FEEL), has recently been pushed into the spotlight, which distributes the task of model training to edge devices and keeps the data locally at the edge devices so as to avoid data uploading and thus preserve user-privacy \cite{Chen2020TWC-joint,Zhu2019TWC_broadband,Liu2020JSAC_privacy}.
Specifically, a typical training process of FEEL involves multiple rounds of wireless communication between the edge server and devices. In a particular round, the edge server firstly broadcasts the global model under training to the edge devices for local \textit{stochastic gradient decent} (SGD) execution using local data, and then the edge devices upload their local models/gradients to the edge server for aggregation and global model updating.
After the convergence criterion is met, such as attaining a desired level of accuracy or reaching a pre-defined value of the loss, the entire training process will be completed, and then on-device models can be tweaked for the edge device's personalization.

In edge networks, the computation resource of the edge devices are constrained, and the wireless resource of the network, e.g., frequency bandwidth, is also limited, so training a AI model by FEEL is usually a time-consuming and expensive task that can take anywhere from hours to weeks to complete \cite{Chen2020TWC_convergence_time}. Hence, training time\footnote{This is also called wall-clock time in some literature \cite{Peter2019arXiv_open,Nori2021TC_tradeoff}. In this work, we use ``total training time'', ``training time'', and ``wall-clock time'' interchangeably.} minimization of AI models is one of the critical concerns in FEEL. The whole training process of FEEL typically consists of multiple communication rounds, and each lasts for a period of time consisting of computation time and communication time, which we called \textit{per-round latency}. To reduce the total training time, we should not only bring down the number of communication rounds, i.e., speed up the convergence rate of the learning algorithm, but also shorten the per-round latency.
Communication-efficient transmission achieved through compression is usually included into the FEEL pipeline to alleviate the transmission burden of the edge devices, and thus reduce the per-round latency \cite{Park2021Proceed_com_efficient}. 
Two main lossy compression techniques, namely quantization and sparsification, as well as combination of them, have been considered in the literature \cite{Amiri2020TSP,Amiri2020TWC,Basu2020JSAIT,Alistarh2017QSGD,Razaviyayn2014SCA,Zhu2020TWC-1bit,Stich2018Sparsified,Wangni2017sparsification,Zhang2021junshan}.
Specifically, in the case of quantization, the gradient vector entries are transmitted after being quantized to finite bits, instead of the full floating-point values \cite{Alistarh2017QSGD,Razaviyayn2014SCA,Zhu2020TWC-1bit}; sparsification reduces the communication overhead by only sending significant entries of the gradient vector \cite{Stich2018Sparsified,Wangni2017sparsification,Zhang2021junshan}.
Although the compressed transmission can decrease the per-round latency, the lossy compression will degrade the convergence speed of FEEL as well, which leads to the increase in the number of communication rounds to achieve a certain accuracy on a given task \cite{Basu2020JSAIT}.
As a result, the compression level balances the \textbf{trade-off between the number of communication rounds and per-round latency} in minimization of the total training time in FEEL.
Besides, the edge devices in the edge network are usually heterogeneous such that some of edge devices with lower computation power become laggards in the synchronous model/gradient aggregation due to their longer computation time, which increases the per-round latency. 
It is necessary to consider the optimization of wireless resources over different edge devices to reduce the communication time of the lagging edge devices, and compensate the longer computation time \cite{Nguyen2020FEEL-IoT,Dinh2020FL-RA,Wan2021arXiv_tradeoff}.
With the goal of minimizing the total training time, we study the following question: How to balance the trade-off between the number of communication rounds and per-round latency via joint quantization level and bandwidth allocation optimization in the presence of device heterogeneity.

\subsection{Related work}

Recently, extensive efforts have been made to minimize the total training time of FEEL by resource allocation or gradient/model compression, which mainly fall into three categories in terms of their objectives:

\subsubsection{Minimization of the number of communication rounds} This is equivalent to accelerating the convergence speed of FEEL algorithms.
In \cite{Chen2020TWC-joint}, the authors minimized the global loss function by optimizing communication resource, e.g., power and bandwidth, and computation resource under given per-round latency constraint, and thus the convergence speed was maximized. 
The works in \cite{Wang2021QuanOutage} and \cite{Salehi2021TC_Unreliable} considered quantization and bandwidth optimization to accelerate the algorithm convergence speed of FEEL with device sampling in the presence of outage probability.
In \cite{Chang2020FL_MAC}, a stochastic gradient quantization was adopted to compress the local gradient, and the quantization levels of each devices were optimized to minimize the optimality gap under multiple access channel capacity constraints. 
These efforts have focused on speeding up the convergence of FEEL algorithms, i.e., reducing the number of communication rounds, without considering minimizing the total training time, which is a more practical and important issue in FEEL.

\subsubsection{Minimization of per-round latency}
The work in \cite{Nguyen2020FEEL-IoT} and \cite{Dinh2020FL-RA} studied the trade-off between per-round latency and energy consumption by introducing a weight factor, and these two objectives tend to form a competitive interaction. 
In \cite{Zhu2019TWC_broadband}, the authors analyzed the per-round latency of different multiple access schemes in FEEL, i.e., the proposed broadband analog aggregation (BAA) and the traditional OFDMA, and proved that the proposed BAA can significantly reduce the per-round latency compared to the traditional OFDMA.
The resource allocation in \cite{Nguyen2020FEEL-IoT}, \cite{Dinh2020FL-RA}, and \cite{Zhu2019TWC_broadband} were considered over each single communication round; nevertheless, FEEL is a long-term process consisting of many communication rounds that together  determine the total training time. 

\subsubsection{Minimization of total training time}
The work in \cite{Wan2021arXiv_tradeoff} minimized the total training time by optimizing communication and computation resource allocation; however, no compression was considered in \cite{Wan2021arXiv_tradeoff}.
The authors in \cite{Chen2020TWC_convergence_time} minimized the training time for a fixed communication rounds by solving a joint learning, wireless resource allocation, and device selection problem. 
Some of other works didn't minimize the training time directly, but they considered to minimize the loss function in a given training time. For example, the work in \cite{Nori2021TC_tradeoff} studied the communication trade-off balanced by compression and local update steps in fixed training time, but the communication resource allocation was not taken into consideration. The authors in \cite{Jin2020one_bit} adopted the idea of \textsc{sign}SGD with majority vote \cite{Bernstein2018signSGD} and optimized the power allocations and CPU frequencies under the trade-off between the number of communication rounds versus the outage probability per communication round for fixed training time.

Despite the above research efforts, these prior works have overlooked the inherent trade-off in minimizing the total training time of communication-efficient FEEL between the number of communication rounds and per-round latency, which is balanced by the quantization level at the edge devices. Moreover, the communication resource allocation among different edge devices and the compression setup for minimizing the total training time of communication-efficient FEEL should be jointly considered. This thus motivates the current work.

\subsection{Our contribution}

This paper studies a FEEL system consisting of multiple edge devices with heterogeneous computational capabilities and one edge server for coordinating the learning process. We consider \textit{quantized FEEL} where a stochastic quantization scheme is adopted for updated gradients compression, which can save per-round communication cost but may at a cost of increased number of communication rounds. Thus, we make a comprehensive analysis on the total training time by taking into account the communication time, computation time and the number of communication rounds, based on which the intrinsic trade-off between the number of communication rounds and per-round latency is characterized. Then, building on the analytical results, a joint quantization and bandwidth allocation optimization problem is formulated and solved. The main contributions are elaborated as follows. 
\begin{itemize}
	\item \textbf{Training time analysis in quantized FEEL:} The challenge of analyzing the total training time mainly lies in estimating the required communication rounds for model convergence. To tackle the challenge, we analyze the convergence behavior of quantized FEEL in terms of the optimality gap and establish the expression of the minimum number of communication rounds that achieves a given optimality gap of the loss function. However, the derived results are generally too loose to be used for further optimization. In order to yield an accurate estimate on the required number of communication rounds, we propose a joint data-and-model-driven fitting method to further refine and tighten the derived result. Thanks to the refinement, an accurate estimate of the total training time can be attained and the trade-off between the number of communication rounds and per-round latency can be better characterized. 
	\item \textbf{Training time minimization via joint optimization of quantization and bandwidth:} Next, building on the derived analytical results, we formulate the total training time minimization problem by jointly optimizing the quantization level and bandwidth allocation, subject to a maximum bandwidth constraint in FEEL network. The problem is non-convex, and thus challenging to solve. To tackle the challenge, we adopt the alternating optimization technique to decompose the problem into two subproblems, and each optimizes one of the two control variables with the other fixed. For the sub-problem of bandwidth allocation with fixed quantization level, it can be solved by bisection search efficiently; for the sub-problem of quantization optimization with bandwidth allocation fixed, a algorithm based on successive convex approximation is proposed.
	\item \textbf{Performance evaluation:} Finally, we conduct extensive simulations to evaluate the performance of task-oriented resource allocation for quantized FEEL by considering the logistic regression (convex loss function) on a synthetic dataset and a convolution neural network (non-convex loss function) on CIFAR-10 dataset. It is shown that the proposed joint data-and-model-driven fitting method can fit the curve of actual optimality gap well. In addition, it is shown that the optimal quantization level found by solving the formulated optimization problem matches well with the simulation result. The benefits of optimizing the bandwidth allocation in coping with the devices heterogeneity are also demonstrated.
\end{itemize}

\subsection{Organization and notations}

\textit{Organization}: The remainder of the paper is organized as follows. Section \ref{sec:system_model} introduces the system model, including the FEEL procedure, the quantization scheme, and the channel models. Section \ref{sec:train_time_analysis} analyzes the convergence of quantized FEEL,where a joint-data-and-model driven fitting approach is proposed to attain a tight approximation of the total training time. The optimization problem for minimizing the total training time is formulated and then solved in Section \ref{sec:time_optimization}. Section \ref{sec:experiment} shows the experimental results using synthetic and real dataset, followed by conclusion in Section \ref{sec:conclusion}.

\textit{Notations}: $\mathbb{R}$ represents the set of real number. $[K]$ denotes set $\{1,2,\ldots,K\}$. $\emptyset$ denotes the empty set. $\mathrm{sgn}(\cdot)$ denotes the sign of a scalar. $\mathbf{w}^{T}$ is transpose of vector $\mathbf{w}$. $\nabla f(\mathbf{w})$ denotes the gradient of function $f$ at point $\mathbf{w}$. $\Vert \mathbf{w} \Vert$ denotes $\ell_{2}$ norm of vector $\mathbf{w}$. $\left\lceil x \right\rceil$ is the ceiling operator. $x\sim\mathcal{CN}(0,\sigma^{2})$ denotes zero-mean circularly symmetric complex Gaussian (CSCG) random variable with variance of $\sigma^{2}$.



\section{System model}\label{sec:system_model}

\subsection{Federated learning}

\begin{figure}[!t]
	\centering
	\includegraphics[width=0.48\textwidth]{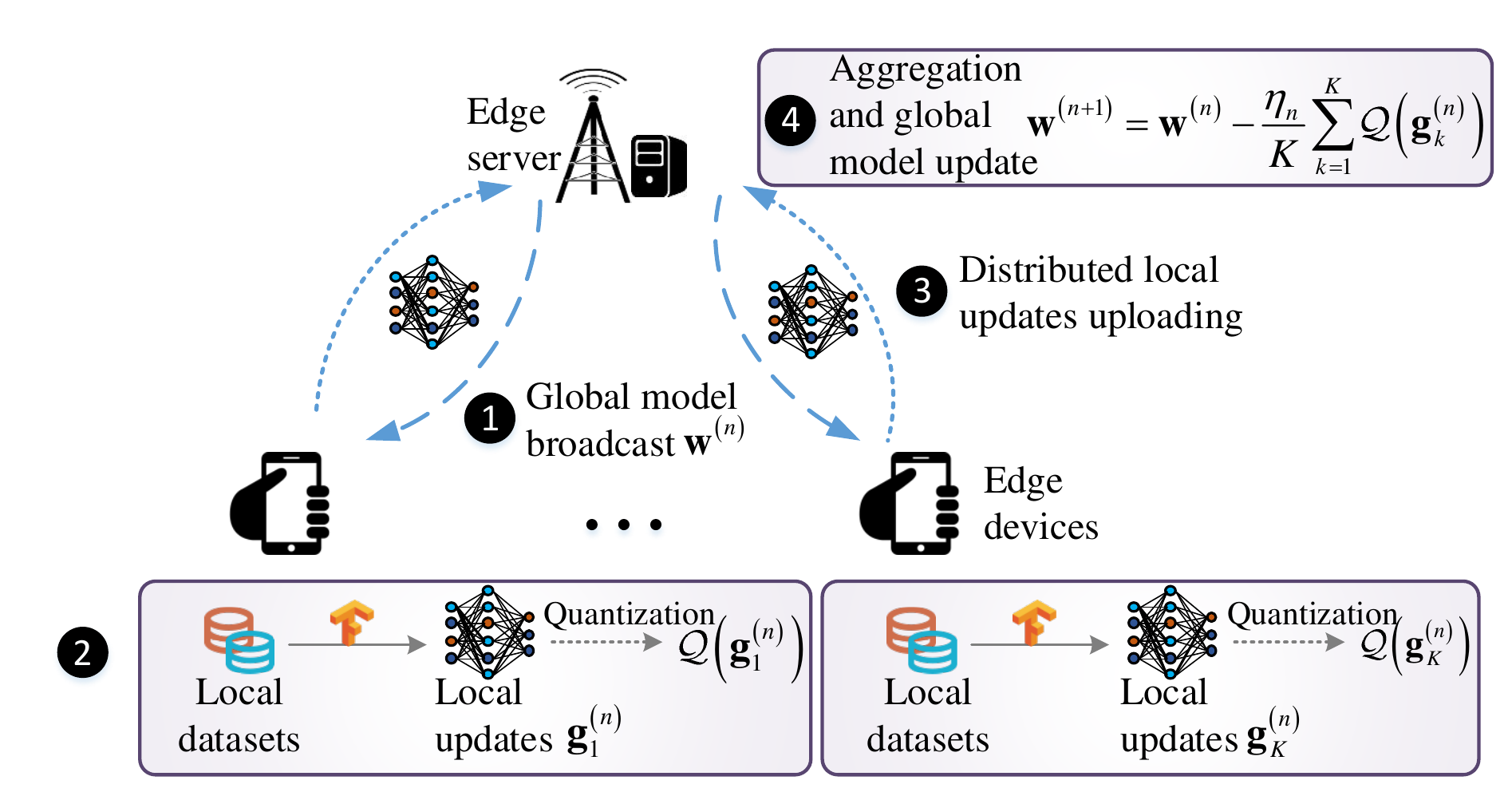}
	\caption{Quantized federated edge learning system with $K$ edge devices.}
	\label{fig:fig-system-model}
\end{figure}

We consider a quantized federated edge learning (FEEL) system consisting of $K$ edge devices and a single edge server, as shown in Fig. \ref{fig:fig-system-model}. With the coordination of the edge server, the edge devices collaboratively train a shared model, which is represented by the parameter vector $\mathbf{w}\in\mathbb{R}^{d}$, with $d$ denoting the model size. The training process is to minimize the following empirical loss function
\begin{equation*}
	F(\mathbf{w}) = \frac{1}{K}\sum_{k=1}^{K}F_{k}(\mathbf{w}),
\end{equation*}
where $F_{k}(\mathbf{w})$ denotes the local loss function at edge device $k$, $k\in [K]$. Suppose device $k$ holds the training data set $\mathcal{D}_{k}$ with a uniform size of $D$, i.e., $\left|\mathcal{D}_{k}\right|=D$. The local loss function $F_{k}(\mathbf{w})$ is given by
\begin{equation}\label{eq:local_loss_function}
	F_{k}(\mathbf{w}) = \frac{1}{D}\sum_{(\mathbf{x}_{i},y_{i})\in\mathcal{D}_{k}}f(\mathbf{w};\mathbf{x}_{i},y_{i}) + \lambda R(\mathbf{w}),
\end{equation}
where $f(\mathbf{w};\mathbf{x}_{i},y_{i})$ denotes the sample-wise loss function specified by the learning task and quantifies the training loss of the model $\mathbf{w}$ on the training data $\mathbf{x}$ and its ground-true label $y_{i}$, and $R(\mathbf{w})$ denotes certain strongly convex regularization function whose strength is controlled by a hyperparameter $\lambda \ge 0$.

In FEEL, the training process is implemented iteratively in a distributed manner using the federated stochastic gradient decent (FedSGD) algorithm as elaborated in the following. Consider a particular iteration or communication round $n$, all the devices first download the current model $\mathbf{w}^{(n)}$ from the server. Then, each device computes a local stochastic gradient, $\mathbf{g}_{k}^{(n)}$, using a randomly chosen mini-batch of samples from data sets $\mathcal{D}_{k}$ in a uniform manner. We denote the set of mini-batch samples used by device $k$ at round $n$ as $\tilde{\mathcal{D}}^{(n)}_{k}$ and the size of each mini-batch as $m_{b}$. Then we have
\begin{equation}\label{eq:gradient}
	\mathbf{g}_{k}^{(n)} = \frac{1}{m_{b}}\sum_{(\mathbf{x}_{i},y_{i})\in\tilde{\mathcal{D}}^{(n)}_{k}}\nabla f(\mathbf{w}^{(n)};\mathbf{x}_{i},y_{i}) + \lambda \nabla R(\mathbf{w}^{(n)}).
\end{equation}
Next, each device transmits a quantized version of its local gradient, i.e., $\mathcal{Q}(\mathbf{g}_{k}^{(n)})$, to the edge server. The quantization scheme will be elaborated in the next subsection. Upon the reception, the edge server aggregates the local gradients and updates the global model as follows:
\begin{equation*}
	\mathbf{w}^{(n+1)} = \mathbf{w}^{(n)} - \frac{\eta_{n}}{K}\sum_{k=1}^{K}\mathcal{Q}(\mathbf{g}_{k}^{(n)}),
\end{equation*}
where $\eta_{n}$ denotes the learning rate at round $n$. Then the updated global model will be broadcast back to all edge devices for initializing the next round training. The above procedure continues until the convergence criterion is met.

\subsection{Stochastic quantization}

We consider a widely-used stochastic quantizer for local gradient quantization \cite{Alistarh2017QSGD}. For any arbitrary vector $\mathbf{g}\in \mathbb{R}^{d}$, the stochastic quantizer $\mathcal{Q}(\mathbf{g}):\mathbb{R}^{d}\rightarrow\mathbb{R}^{d}$ is defined element-wisely as 
\begin{equation}
	{\cal Q}\left( {{g_i}} \right) = \left\| {\mathbf{g}} \right\| \cdot {\mathop{\rm sgn}} \left( {{g_i}} \right) \cdot {\xi _i}\left( {{\mathbf{g}},q} \right),\;\forall i \in \left[ d \right],
\end{equation}
where the output of quantizer $\mathcal{Q}(\mathbf{g})$ consists of three parts, i.e., the vector norm $\Vert\mathbf{g}\Vert$, the sign of each entry $\mathrm{sgn}(g_{i})$ with $g_{i}$ denoting the $i$-th entry of $\mathbf{g}$, and the quantization value of each entry $\xi_{i}(\mathbf{g},q)$. $\left\lbrace{\xi _i}\left( {{\mathbf{g}},q} \right)\right\rbrace$ are independent random variables defined as
\begin{equation*}
	{\xi _i}\left( {{\bf{g}},q} \right) = \left\{ {\begin{array}{*{20}{c}}
			(l+1)/q,&{\text{w.p.}\;\frac{{\left| {{g_i}} \right|}}{{\left\| {\bf{g}} \right\|}}q - l},\\
			l/q,&{{\text{otherwise.}}}
	\end{array}} \right.
\end{equation*}
Here, $q$ denotes the the number of quantization levels and $0\leq l < q$ is an integer such that $\frac{g_i}{\left\| {\bf{g}} \right\|} \in \left[ \frac{l}{q},\frac{l+1}{q}\right)$.

As proved in Lemma 3.1 from \cite{Alistarh2017QSGD}, the random quantizer $\mathcal{Q}(\mathbf{g})$ is unbiased, i.e., $\mathbb{E}[\mathcal{Q}(\mathbf{g})]=\mathbf{g}$ for any given vector $\mathbf{g}$. Moreover, assuming that $d \ge q^2$, the quantizer has a bounded variance, i.e., $\mathbb{E}[\left\Vert\mathcal{Q}(\mathbf{g})-\mathbf{g}\right\Vert]^{2} \leq \frac{\sqrt{d}}{q}\left\Vert\mathbf{g}\right\Vert^{2}$. Note that we do not claim that this quantization scheme is optimal in terms of the communication efficiency. Rather, we adopt it as a simple and general scheme that facilitates the subsequent analysis of the trade-off between the number of communication rounds and per-round communication latency, controlled by the quantization level.

\subsection{Wireless transmission model}\label{sec:transmission-model}

In the quantized FEEL system, each edge device connects to the edge server via a shared wireless medium. We assume that the spectrum is divided into distinct and non-overlapping flat fading channels with different bandwidth, so that the edge devices share the spectrum through frequency division multiple access to avoid interferences with each other. 
In general, modern neural network models are of high dimension with $d$ in the order of $10^6 \sim 10^9$. Hence, it usually takes much longer than the coherence period to transmit a complete model. For example, a single LTE frame of 5 MHz bandwidth and 10 ms duration can carry only 6000 complex symbols \cite{Amiri2020TWC}. To transmit a moderate neural network model with $10^6$ parameters encoded by 32-bit floating-point values, it will approximately take 6 seconds, which is much longer than the frame length, i.e., 10 ms. Moreover, in some Internet of Things (IoT) networks, which are typically bandwidth and power limited \cite{Dhillon2017Bandwidth}, it takes more time to transmit a machine learning model. Hence, it is reasonable to model the wireless uplink channels as independent and identical distributed (i.i.d.) \textit{fast Rayleigh fading channel} in the course of training, i.e., the channel coefficients remain constant over each coherence period and vary in an i.i.d. across different coherence periods, and the codeword or frame will span multiple coherence periods \cite{Tse2005Fundamentals}. Specifically, the channel propagation coefficient between edge server and device $k$ is generally modeled as $h_{k} = \sqrt{\phi_{k}}\overline{h}_{k}$; here, $\phi_{k}$ describes the large-scale propagation effects, including path loss and shadowing, and $\overline{h}_{k}$ denotes small-scale fading modeled as i.i.d. CSCG random variables with zero mean and unit variance, i.e., $\overline{h}_{k} \sim \mathcal{CN}(0,1)$. The large-scale propagation coefficient $\phi_{k}$ remains unchanged in a whole time frame, while the small-scale fading $\overline{h}_{k}$ varies from coherence block to another in a time frame. Moreover, we assume that the channel coefficients are only known at the edge server, which can be obtained by channel estimation at the server.

In this situation, the ergodic capacity can be assigned to the fast fading channel, which can be achieved in practice by interleaving technique \cite{Tse2005Fundamentals}. The ergodic capacity of device $k$ is given by
\begin{equation}\label{eq:ergordic_rate}
	R_{k}=\mathbb{E}_{h_{k}}\left[b_{k}\log_{2}\left(1+\frac{p_{k}\vert h_{k} \vert^{2}}{b_{k}N_{0}}\right)\right],
\end{equation}
where $b_{k}$ denotes the frequency bandwidth allocated for device $k$ with $\sum_{k=1}^{K}b_{k}=B_{0}$; $p_{k}$ denotes the transmit power at device $k$; $N_{0}$ denotes noise power spectral density; the expectation is taken over the channel distribution. 

\section{Training time analysis}\label{sec:train_time_analysis}

\begin{figure}[!t]
	\centering
	\includegraphics[width=0.48\textwidth]{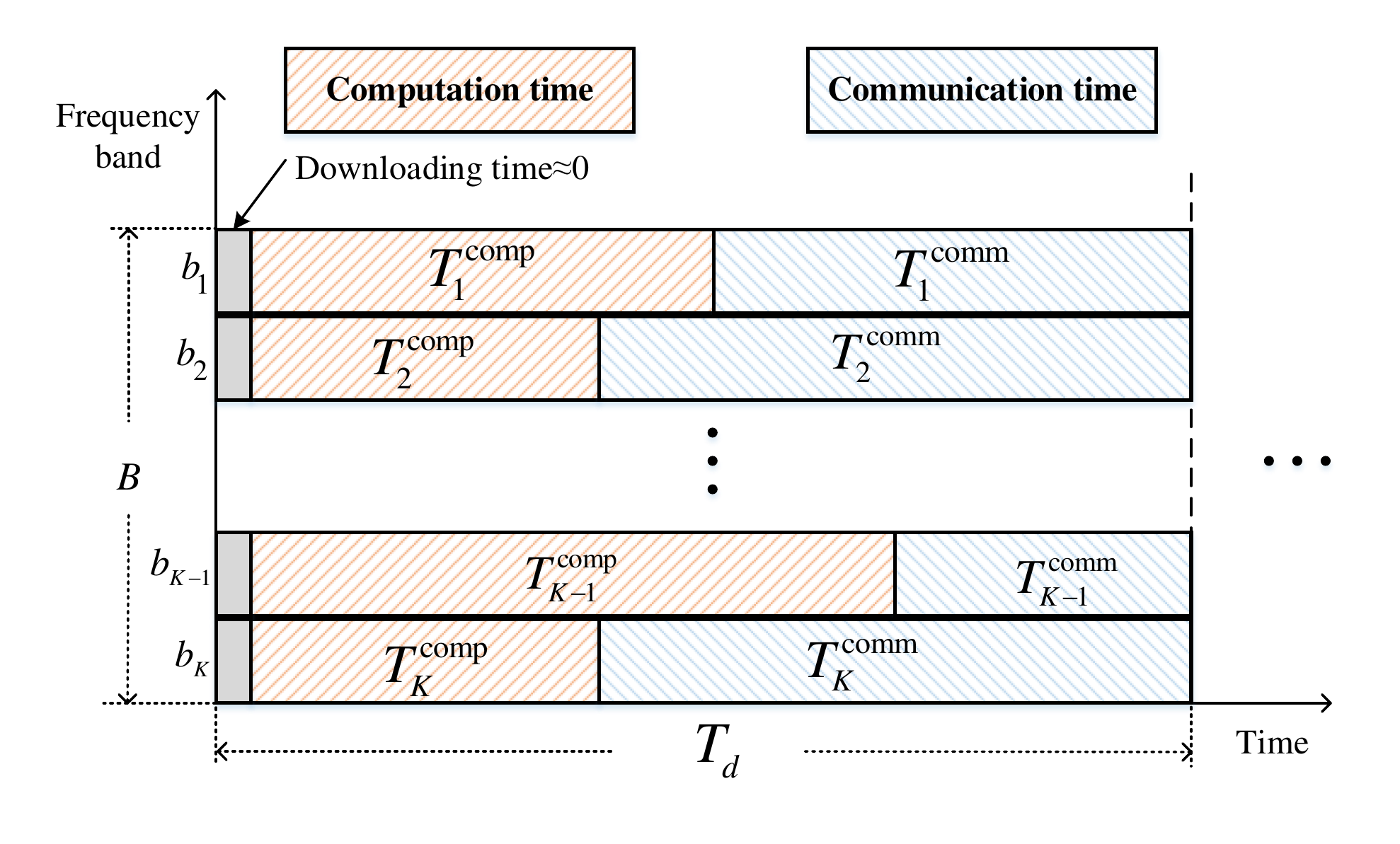}
	\caption{Computation time and communication time in one communication round of quantized FEEL.}
	\label{fig:fig-time}
\end{figure}

The training time of each device for one communication round comprises computation time $T_{k}^{\text{comp}}$ and communication time $T_{k}^{\text{comm}}$, as illustrated in Fig. \ref{fig:fig-time}. Since the server broadcasts the same global model to all the devices using the entire frequency band, the downlink delay due to global model broadcasting is ignorable compared with the uplink delay due to updates uploading from many devices to the server. Assume that the delay requirement for running one round training is $T_{d}$, i.e., $T_{k}^{\text{comp}}+T_{k}^{\text{comm}} \leq T_{d}$. We define $N_{\epsilon}$ as the minimum number of communication rounds when $\epsilon$-optimality gap is achieved, i.e., $F(\mathbf{w}^{(N_{\epsilon})})-F(\mathbf{w}_{\ast}) \leq \epsilon$, where $\mathbf{w}_{\ast}$ denotes the optimal model as $\mathbf{w}_{\ast} = \arg\min\limits_{\mathbf{w}}F(\mathbf{w})$. The training process stops when $\epsilon$-optimality gap is achieved.
Then, the requirement of total training time over $N_{\epsilon}$ rounds is given by
\begin{equation}\label{eq:total-time}
	T = N_{\epsilon} \cdot T_{d}.
\end{equation}
In the following, we give the expression of per-round training time $T_d$, and obtain the minimum number of communication rounds $N_{\epsilon}$ by analyzing the convergence of FEEL, which paves the way for minimizing the total training time defined in (\ref{eq:total-time}) in next section.

\subsection{Computation time}

Let $\nu$ denote the number of processing cycles for one particular edge device to execute one batch of samples, and $f_{k}$ denote the CPU frequency of device $k$. Accordingly, the computation time for running one-round SGD at device $k$ is given by \cite{Ren2020TWC}
\begin{equation*}
	T_{k}^{\text{comp}} = \frac{\nu}{f_{k}}.
\end{equation*}

\subsection{Communication time}

Let $S$ denote the number of bits for transmission after stochastic quantization. For quantizing any element $g_{i}$ in gradient vector $\mathbf{g}\in \mathbb{R}^{d}$ as noted from (\ref{eq:gradient}), we need to encode the vector norm $\Vert\mathbf{g}\Vert$, the element-wise sign $\mathrm{sgn}(g_{i})$, and the normalized quantization value $\xi_{i}(\mathbf{g},q)$ into bits. Particularly, it takes one bit to encode each of the $\mathrm{sgn}(g_{i})$. Assuming that $\xi_{i}(\mathbf{g},q)$'s have uniform distribution\footnote{In practice, $\xi_{i}(\mathbf{g},q)$'s can have non-uniform probability distribution. There exit more efficient coding methods coping with such non-uniform distribution as reported in \cite{Alistarh2017QSGD}, which are nevertheless much more complicated for analysis and optimization. In this work, we only consider the uniform distribution and leave those complicated ones for future work.} with support $\{0, 1/q, 2/q, \dots, 1$\}, it takes at least $\log_2(1+q)$ bits to encode each $\xi_{i}(\mathbf{g},q)$ \cite{Cover2006IT}. Since each vector containing $d$ entries, it takes in total $(1+\log_{2}q)d$ bits to encode these two parts. By contrast, the overhead in encoding the single scalar vector norm $\Vert\mathbf{g}\Vert$ is typically negligible for large models of size $d$ \cite{Shlezinger2020UVeQFed}. To facilitate the subsequent analysis, for large $d$ we approximate $S$ via
\begin{equation}\label{eq:bits_in_total}
	S = (1+\log_{2}(q+1))d.
\end{equation}
The communication delay in one round is 
\begin{equation}\label{eq:communication_time}
	T_{k}^{\text{comm}} = \frac{S}{R_{k}},
\end{equation}
where $R_{k}$ is the ergodic capacity defined in (\ref{eq:ergordic_rate}).

\subsection{Minimum number of communication rounds}\label{sec:commun-rounds}

In this subsection, we derive $N_{\epsilon}$ by analyzing the convergence of quantized FEEL. To this end, we make the following assumptions on the local loss functions $\{F_{k}(\mathbf{w})\}$.

\begin{assumption}[Smoothness]\label{asp:smoothness}
	The local loss functions $\{F_{k}(\mathbf{w})\}$ are all $L$-smooth: for all $\mathbf{w}_{i}$ and $\mathbf{w}_{j}$, $F_{k}(\mathbf{w}_{i}) \leq F_{k}(\mathbf{w}_{j}) + \left(\mathbf{w}_{i} - \mathbf{w}_{j}\right)^{T}\nabla F_{k}(\mathbf{w}_{j}) + \frac{L}{2}\left\Vert \mathbf{w}_{i} - \mathbf{w}_{j} \right\Vert^{2}$, $\forall k$.
\end{assumption}

\begin{assumption}[Strongly convexity]\label{asp:convexity}
	The local loss functions $\{F_{k}(\mathbf{w})\}$ are all $\mu$-strongly convex: for all $\mathbf{w}_{i}$ and $\mathbf{w}_{j}$, $F_{k}(\mathbf{w}_{i}) \ge F_{k}(\mathbf{w}_{j}) + \left(\mathbf{w}_{i} - \mathbf{w}_{j}\right) + \left(\mathbf{w}_{i} - \mathbf{w}_{j}\right)^{T}\nabla F_{k}(\mathbf{w}_{j}) + \frac{\mu}{2}\left\Vert \mathbf{w}_{i} - \mathbf{w}_{j} \right\Vert^{2}$, $\forall k$.
\end{assumption}

\begin{assumption}[First and second moments of local gradients]\label{asp:gradient}
	The mean and variance of stochastic gradients $\mathbf{g}_{k}^{(n)}$ of local loss functions $F_{k}(\mathbf{w})$, for all $n\in [N]$ and $\forall k$, satisfy that
	\begin{align*}
		\text{(Unbiased)}\;\;\;&\mathbb{E}[\mathbf{g}_{k}^{(n)}] = \nabla F_{k}(\mathbf{w}^{(n)}),\\
		\text{(Bounded variance)}\;\;\;&\mathbb{E}[\Vert\mathbf{g}_{k}^{(n)} - \nabla F_{k}(\mathbf{w}^{(n)})\Vert^{2}] \leq \delta_{k}^{2}.
	\end{align*}
\end{assumption}

Assumptions \ref{asp:smoothness} and \ref{asp:convexity} on local loss functions are standard, and they can be satisfied by many typical learning models, such as logistic regression, linear regression, and softmax classifier. Assumption \ref{asp:gradient} is general enough to cope with both i.i.d. and non-i.i.d. data distribution across edge devices, which follows the work in \cite{Li2020ICML}, \cite{Salehi2021TC_Unreliable}, and \cite{Luo2020cost}.
Under Assumptions \ref{asp:smoothness}-\ref{asp:gradient}, the convergence rate of quantized FEEL is established in the following theorem.

\begin{theorem}\label{theorem:convergence}
	Consider a quantized FEEL system with fixed quantization level $q \ge 2$.  The optimality gap of the loss function after $N$ communication rounds is upper bounded by
	\begin{align*}
		&\mathbb{E}\left[F(\mathbf{w}^{(N)})\right]-F(\mathbf{w}_{\ast})\\
		&\leq \frac{\alpha \kappa}{N+2\alpha \kappa-1}\left(L\left\Vert \mathbf{w}^{(0)} - \mathbf{w}_{\ast} \right\Vert^2 + \frac{2\Gamma}{\mu}\right),
	\end{align*}
	where $\alpha = \frac{\sqrt{d}}{qK}+1$, $\kappa = \frac{L}{\mu}$, $F_{\delta} = F(\mathbf{w}_{\ast}) - \frac{1}{K}\sum_{k=1}^{K}F_{k}^{\ast}$ with $F_{k}^{\ast}=\min\limits_{\mathbf{w}}F_{k}(\mathbf{w})$, $\Gamma = 2 LF_{\delta} +\frac{1}{K}\sum_{k=1}^{K}\delta_{k}^{2}$, and $\mathbf{w}^{(0)}$ is the initial point of the training process. The learning rate is set to be a diminishing one, i.e., $\eta_{n}=\frac{2}{\mu(n+2 \alpha\kappa-1)}$.
\end{theorem}
\begin{proof}
	See Appendix \ref{apdx:theorem_1}.
\end{proof}

\begin{remark}[Convergence rate]
	Theorem \ref{theorem:convergence} quantizes the impact of gradient quantization on the convergence rate of the FEEL, which is captured by the term $\alpha=\frac{\sqrt{d}}{qK}+1$. An aggressive quantization scheme, e.g., with small $q$, will lead to a enlarged optimality gap, and thus needs more rounds to converge. Nevertheless, the quantized FEEL can still achieve the asymptotic convergence rate of $\mathcal{O}(\frac{1}{N})$ as federated learning without quantization \cite{Li2020ICML}.
\end{remark}

\begin{remark}[Impact of data heterogeneity]
	The term $F_{\delta}$ measures the discrepancy between the minimum global loss and the average of the minimum local losses, which can be used for quantifying the heterogeneity level of data distribution among different devices \cite{Li2020ICML}. As observed, if the data distribution is i.i.d., i.e., the data at different devices is sampled from a same distribution, then $F_{\delta}$ goes to zero as the number of samples grows; otherwise, $F_{\delta}$ is a non-zero constant depending on the skewness of data distribution. Therefore, Theorem \ref{theorem:convergence} also offers quantitative insights on how data heterogeneity affects the convergence of quantized FEEL.
\end{remark}

Although we can readily establish a bound on $N_{\epsilon}$ based on Theorem \ref{theorem:convergence} and apply this bound for subsequent resource optimization as did in prior work, e.g., \cite{Wang2019JSAC} and \cite{Luo2020cost}, such an approach has two key drawbacks. Firstly, the gap between the derived bound and its true value could be large as several relaxations are made in deriving the bound. Thus, ignoring the gap may lead to highly suboptimal solution in the subsequent resource optimization. Secondly, even we adopt the upper bound and ignore the effects of the gap, it is still difficult to get the exact value of upper bound, since it involves calculating a bunch of data-related and model-related parameters, such as $\mu$, $L$, $F_{\delta}$, $\mathbf{w}_{\ast}$, and $\Gamma$. 

To address the above issues, we propose a joint data-and-model-driven fitting approach, which uses a small amount of pre-training rounds to yield a good estimate of the optimality gap based on the bound derived in Theorem \ref{theorem:convergence}.
To this end, we first denote the upper bound function derived in Theorem \ref{theorem:convergence} as follows, 
\[\hat{U}(N) = \frac{\alpha \kappa \left(L\left\Vert \mathbf{w}_{0} - \mathbf{w}_{\ast} \right\Vert^2 + \frac{2\Gamma}{\mu}\right)}{N+2\alpha \kappa-1}.\] 
Before we derive the tight estimate of optimality gap, it is first observed that the upper bound function $\hat{U}(N)$ satisfies the following properties:
\begin{enumerate}
	\item $\hat{U}(N)$ is a decreasing function of $N$, and it converges to zero in a rate of $\mathcal{O}(\frac{1}{N})$;
	\item $\hat{U}(N)$ has a fractional structure, where the numerator and denominator are both linear increasing functions of $\alpha$.
\end{enumerate}
We assume that the ground-true optimality gap follows the same properties as its upper bound $\hat{U}(N)$. Based on the above assumption, the exact  optimality gap can be well estimated by the following function:
\begin{equation}\label{eq:fitting_expression}
	\mathbb{E}\left[F(\mathbf{w}^{(N)})\right]-F(\mathbf{w}_{\ast}) = \frac{\alpha A + D}{n + \alpha B + C} \triangleq U(N),
\end{equation}
where $A>0$, $B>0$, $C\ge 0$, and $D \ge 0$ are tuning parameters to be fitted, which are implicitly related to the parameters, such as $\mu$, $L$, $F_{\delta}$, and $\Gamma$. It can be seen that $U(N)$ generalizes all the functions that satisfy the two properties mentioned above. 

Next, we apply a joint data-and-model-driven fitting method to fit the values of the tuning parameters as follows. Firstly, we randomly choose two quantization levels, say $q_{1}$ and $q_{2}$, and run the quantized FEEL with $q_{1}$ and $q_{2}$, respectively, from an initial model $\mathbf{w}_{0}$. Then we sample the value of loss at each round until the number of communication rounds reaches a pre-defined value $\tilde{N}$. The corresponding loss values are denoted as $F_{i,n}$ ($i\in\{1,2\}$, $n\in [1,\tilde{N}]$) for round $n$ when the quantization level is $q_i$. According to (\ref{eq:fitting_expression}), we have that
\begin{equation}\label{eq:estimation_upper_bound}
	F_{i, n}-Z \approx \frac{X_{i}}{n + Y_{i}},\;\forall i\in\{1,2\},\;n\in [1,\tilde{N}],
\end{equation}
where $\alpha_{i} = \frac{\sqrt{d}}{q_{i}K}+1$, $Z=F(\mathbf{w}_{\ast})$, and
\begin{empheq}[left=\empheqlbrace]{align}
	\label{eq:numerator} X_{i} &= \alpha_{i} A + D,\\
	\label{eq:denominator} Y_{i} &= \alpha_{i} B + C.
\end{empheq}
Then we aim to find proper values of $X_{i}$, $Y_{i}$, and $Z$ to fit (\ref{eq:estimation_upper_bound}) well. The method we choose is to solve the following nonlinear regression problem:
\begin{equation}\label{eq:problem_xyz}
	\underset{X_{i},Y_{i}, Z}{\text{min}}
	\;\sum_{i=1}^{2}\sum_{n=1}^{\tilde{N}}\big((F_{i, n}-Z)(n + Y_{i})-X_{i}\big)^2.
\end{equation}
For any fixed $Z$, this problem can be divided into two linear regression problems, i.e.,
\begin{equation}
	\underset{X_{i},Y_{i}}{\text{min}}
	\;\;\;\sum_{n=1}^{\tilde{N}}\big((F_{i, n}-Z)(n + Y_{i})-X_{i}\big)^2,
\end{equation}
and the optimal $\{X_{i}\}$ and $\{Y_{i}\}$ can be given by
\begin{equation}\label{eq:Xi}
	X_{i} = \frac{\sum_{n=1}^{\tilde{N}}\chi_{i,n}\sum_{n=1}^{\tilde{N}}\psi_{i,n}^{2}-\sum_{n=1}^{\tilde{N}}\chi_{i,n}\psi_{i,n}\sum_{n=1}^{\tilde{N}}\psi_{i,n}}{N\sum_{n=1}^{\tilde{N}}\psi_{i,n}^2-(\sum_{n=1}^{\tilde{N}}\psi_{i,n})^2},
\end{equation}
and 
\begin{equation}\label{eq:Yi}_{}
	Y_{i} = \frac{\sum_{n=1}^{\tilde{N}}\chi_{i,n}\sum_{n=1}^{\tilde{N}}\psi_{i,n}-N\sum_{n=1}^{\tilde{N}}\chi_{i,n}\psi_{i,n}}{N\sum_{n=1}^{\tilde{N}}\psi_{i,n}^2-(\sum_{n=1}^{\tilde{N}}\psi_{i,n})^2}
\end{equation}
where $\chi_{i,n} = (F_{i,n}-Z)n$ and $\psi_{i,n} = F_{i,n}-Z$. Hence, the problem in (\ref{eq:problem_xyz}) can be solved by one-dimensional search of $Z$. Since the computations in (\ref{eq:Xi}) and (\ref{eq:Yi}) only involve limited algebraic operations, the computation time for solving the problem in (\ref{eq:problem_xyz}) is negligible compared with the whole training process. With $\{X_{i}\}$ and $\{Y_{i}\}$ at hand, $A$, $B$, $C$, $D$ can be obtained from (\ref{eq:numerator}) and (\ref{eq:denominator}) as following:
\begin{empheq}[left=\empheqlbrace]{align*}
	A & = \frac{X_{1}-X_{2}}{\alpha_{1}-\alpha_{2}},\\
	B & = \frac{Y_{1}-Y_{2}}{\alpha_{1}-\alpha_{2}},\\
	C & = \frac{\alpha_{2}Y_{1}-\alpha_{1}Y_{2}}{\alpha_{2}-\alpha_{1}},\\
	D & = \frac{\alpha_{2}X_{1}-\alpha_{1}X_{2}}{\alpha_{2}-\alpha_{1}}.
\end{empheq}

Based on the estimated optimality gap in (\ref{eq:fitting_expression}) with the well-fitted parameters, we can derive $N_{\epsilon}$ as in the following proposition. 

\begin{proposition}\label{coro}
	In the quantized FEEL system, the minimum communication rounds to achieve $\epsilon$-optimality gap is given by
	\begin{equation}\label{eq:T_epsilon}
		N_{\epsilon} = \left\lceil \left(\frac{\sqrt{d}}{qK}+1\right)\left(\frac{A}{\epsilon}-B\right) + \frac{D}{\epsilon}-C \right\rceil.
	\end{equation}
\end{proposition}

\begin{proof}
	By setting the fitted optimality gap in (\ref{eq:fitting_expression}) less than $\epsilon$, i.e., $U(N) \leq \epsilon$, and respecting the fact that the minimum communication rounds should be an integer, we obtain (\ref{eq:T_epsilon}).
\end{proof}

\begin{remark}[Impact of quantization level and device number]\label{remark}
	Proposition \ref{coro} unveils the impact of quantization level and the number of participating devices on the minimum communication rounds as reflected by the term $\frac{\sqrt{d}}{qK}+1$. On one hand, $N_{\epsilon}$ decreases with increasing number of the quantization level $q$. This is due to the fact that increasing the quantization levels leads to less quantization error, which speeds up the convergence. On the other hand, we can observe that as the number of devices goes to infinity, i.e., $K \rightarrow \infty$, the impact of quantization diminish since the quantization errors average out thanks to the update aggregation mechanism. Moreover, We can obtain that $N_{\epsilon} =  \left\lceil \frac{A+D}{\epsilon}-B -C \right\rceil$ when $q \rightarrow \infty$ or $K \rightarrow \infty$. In other words, $\left\lceil \frac{A+D}{\epsilon}-B -C \right\rceil$ can be used to evaluate the minimum communication rounds under high-resolution quantization or sufficient large number of devices, and this value offers us a lower bound of the minimum communication rounds under practical quantization levels and number of devices.
\end{remark}

\section{Training time minimization}\label{sec:time_optimization}

In this section, we aim to jointly optimize the quantization level and the bandwidth allocation by minimizing the training time defined in (\ref{eq:total-time}) to achieve an $\epsilon$-optimality gap. 

\subsection{Problem formulation}

The training time minimization problem is mathematically formulated by
\begin{align}
	\label{eq:problem}(\text{P1})\;\;\;\underset{q\in\mathbb{Z}^{+},\{b_{k}\},T_{d}}{\text{min   }}
	\;\;\;&T_{d}\cdot N_{\epsilon},\\  \text{s.t.} \;\;\;
	&\nonumber T_{k}^{\text{comp}}+T_{k}^{\text{comm}} \leq T_{d},\;\forall k\in[K],\tag{\ref{eq:problem}a}\\
	&\nonumber \sum_{k=1}^{K}b_{k} = B_{0},\tag{\ref{eq:problem}b}\\
	&\nonumber q \ge 2,\tag{\ref{eq:problem}c}
\end{align}
where the objective function in (\ref{eq:problem}) is the total training time needed to achieve $\epsilon$-optimality gap. Constraints in (\ref{eq:problem}a) indicate that the training time of each device per communication round cannot exceed the delay requirement $T_{d}$. Equation (\ref{eq:problem}b) constrains total bandwidth allocated to all the devices as $B$. The constraint on the quantization level $q$ is described by (\ref{eq:problem}c).

The objective function in (\ref{eq:problem}) is complicated due to the coupling of the control variables $T_{d}$ and $q$. Moreover, $q$ can only take values from positive integers. Therefore, Problem (P1) is non-convex, and challenging to be solved optimally. To yield a good solution to Problem (P1), we divide it into two sub-problems. One is finding the optimal bandwidth allocation $\{b_{k}\}$ and $T_{d}$ with fixed quantization level $q$; the other is finding the optimal quantization level $q$ with fixed bandwidth allocation $\{b_{k}\}$ and $T_{d}$. We will find that the first sub-problem can be solved optimally and efficiently with unique solution, and the second sub-problem can be transformed into a non-convex problem that can be solved by the method of successive convex approximation (SCA) \cite{Razaviyayn2014SCA}. Then, by alternatively solving each sub-problem, we can obtain good sub-optimal solutions to joint quantization level and bandwidth allocation optimization.

\begin{remark}[Tradeoff between minimum communication rounds and per-round latency]
	As noted in Remark \ref{remark}, the minimum communication rounds $N_{\epsilon}$ can be reduced by increasing the quantization levels $q$, but at a cost of increased per-round latency. Therefore, there exists a fundamental trade-off between reducing the minimum communication rounds and suppressing the per-round latency when minimizing the total training time. The trade-off is manipulated by the setting of quantization level $q$. 
\end{remark}

\begin{remark}[Resource allocation over heterogeneous devices in FEEL]
	The computation time of the devices varies due to their heterogeneous computation capacity. To enforce the per-round latency constraint, more frequency bandwidth should be allocated to the devices with low computation power so as to compensate the long computation time with short communication time, and vice versa. Hence, the bandwidth allocation among the devices should jointly account for the channel condition and also the computation resources, which is in sharp contrast to classic bandwidth allocation problem account for only the channel condition, e.g., in \cite{Gong2010TSP}.
\end{remark}

\subsection{Bandwidth allocation optimization}

Since $N_{\epsilon}$ is independent with $\{b_{k}\}$ and $T_{d}$,  Problem (\ref{eq:problem}) under fixed quantization level $q$ reduces to 
\begin{align}
	\label{eq:problem_band}\text{(P2)}\;\;\;\underset{\{b_{k}\},T_{d}}{\text{min   }}
	\;\;\;&T_{d}\\  \text{s.t.} \;\;\;
	&\nonumber T_{k}^{\text{comp}}+T_{k}^{\text{comm}} \leq T_{d},\;\forall k\in[K],\tag{\ref{eq:problem_band}a}\\
	&\nonumber \sum_{k=1}^{K}b_{k} = B_{0}.\tag{\ref{eq:problem_band}b}
\end{align}

Since $T_{k}^{\text{comm}}=\frac{S}{R_{k}}$ as defined in (\ref{eq:communication_time}), constraint (\ref{eq:problem_band}a) can be rewritten as 
\begin{equation*}
	T_{k}^{\text{comp}}+\frac{S}{R_{k}} \leq T_{d},\;\forall k\in[K].
\end{equation*}
To get a closed-form expression of $R_{k}$, it can be rewritten as
\begin{align*}
	\nonumber R_{k} &= \int_{0}^{+\infty}b_{k}\log_{2}\left(1+\frac{p_{k}x}{b_{k}N_{0}}\right)f_{\vert h_{k} \vert^{2}}(x)dx\\
	\label{eq:R_CDF}&=\frac{b_{k}}{\ln 2}\frac{p_{k}}{b_{k}N_{0}}\int_{0}^{+\infty}\frac{1-F_{\vert h_{k} \vert^{2}}(x)}{1+\frac{p_{k}x}{b_{k}N_{0}}}dx,
\end{align*}
where $f_{\vert h_{k} \vert^{2}}(x)$ and $F_{\vert h_{k} \vert^{2}}(x)$ are the probability density function (PDF) and cumulative distribution function (CDF) of the random variable $\vert h_{k} \vert^{2}$, respectively. It can be verified that $\vert h_{k} \vert^{2}$ follows an exponential distribution, i.e., $\vert h_{k} \vert^{2} \sim \text{Exp}(1/\phi_{k})$. Hence, we have $F_{\vert h_{k} \vert^{2}}(x) = 1 - e^{-x/\phi_{k}}$. Then, $R_{k}$ can be calculated as 
\begin{align*}
	R_{k} &= \frac{b_{k}}{\ln 2}\frac{p_{k}}{b_{k}N_{0}}\int_{0}^{+\infty}\frac{e^{-x/\phi_{k}}}{1+\frac{p_{k}x}{b_{k}N_{0}}}dx\\
	&=\frac{b_{k}}{\ln 2}\int_{0}^{+\infty}\frac{e^{-x/\phi_{k}}}{x+\frac{b_{k}N_{0}}{p_{k}}}dx.
\end{align*}
According to \cite[Section 8.212]{Gradshteyn2014Integrals}, we have, for real number $a$ and $b>0$, $\int_{0}^{+\infty}\frac{e^{-bx}}{a+x}dx=-e^{ab}\text{Ei}(-ab)$, where $\text{Ei}(x)=\int_{-\infty}^{x}\frac{e^{\rho}}{\rho}d\rho$ is the exponential integral function, $R_{k}$ can be rewritten in closed-form as follows,
\begin{equation}\label{eq:rate_close_form}
	R_{k} = -\frac{b_{k}}{\ln 2}e^{b_{k}\theta_{k}}\text{Ei}(-b_{k}\theta_{k}),
\end{equation}
where $\theta_{k} = \frac{\underline{}N_{0}}{p_{k}\phi_{k}}$.

It can be verified that the transmission rate $R_{k}$ in (\ref{eq:rate_close_form}) is an increasing function of $b_{k}$, which we denote as $R_{k}(b_{k})$. Hence, $T_{k}^{\text{comm}} = \frac{S}{R_{k}(b_{k})}$ decreases with increasing $b_{k}$. The following lemma will be beneficial for solving Problem (P2).
\begin{lemma}\label{prop}
	Constraints (\ref{eq:problem_band}a) in Problem (P2) can be replaced by 
	\begin{equation*}
		T_{k}^{\text{comp}}+T_{k}^{\text{comm}} = T_{d},\;\forall k\in[K].
	\end{equation*}
\end{lemma}
\begin{proof}
	Please see Appendix \ref{apdx:prop_1}.
\end{proof}

From Lemma \ref{prop}, each $b_{k}$ can be represented as a function of $T_{d}$, i.e.,
\begin{equation}\label{eq:Rk_inverse}
	b_{k}(T_{d}) = R^{-1}_{k}\left(\frac{S}{T_{d}-T_{k}^{\text{comp}}}\right),
\end{equation}
where $R^{-1}_{k}(\cdot)$ denotes the inverse function of $R_{k}(\cdot)$. Since it holds that $\sum_{k=1}^{K}b_{k} = B_{0}$, we can find $T_{d}$ by solving the equation as follows
\begin{equation}\label{eq:solving_td}
	\sum_{k=1}^{K}b_{k}(T_{d}) = B_{0}.
\end{equation}
It can be verified that $b_{k}(T_{d})$ is a decreasing function of $T_{d}$. Therefore, Equation (\ref{eq:solving_td}) can be efficiently solved by bisection search. Note that although $R_{k}(\cdot)$ is an increasing function with closed form expression, it is nontrivial to obtain a tractable expression of $R^{-1}_{k}(\cdot)$, so we can not get $b_{k}$ with given $T_{d}$ directly from (\ref{eq:Rk_inverse}). Instead, $b_{k}$ with given $T_{d}$ can be obtained by solving
\begin{equation*}
	R_{k}(b_{k})=\frac{S}{T_{d} - T_{k}^{\text{comp}}},
\end{equation*}
with the tool of bisection search. This is feasible due to the monotonicity of $R_{k}$. 

In consequence, Problem (P2) can be solved by two-layer bisection search as summarized in \textbf{Algorithm 1}. In the outer layer bisection, we search $T_{d}$ in the range of $[T_{d}^{-},T_{d}^{+}]$, where $T_{d}^{-}=\max\limits_{k}\{T_{k}^{\text{comp}}\}$ and $T_{d}^{+} = \max\limits_{k}\{T_{k}^{\text{comp}}+R_{k}(B_{0}/K)\}$.
The inner layer bisection in Step 5 is implemented in the range of $[0,B_{0}]$. Since only single variable $b_{k}$ involves in each bisection, the process is straightforward, and thus we omit the detailed steps for simplicity.

\begin{algorithm}\small
	\caption{Two-layer bisection search for solving Problem (P2)}
	\label{algo}
	\begin{algorithmic}[1]
		\State{Input parameters: $B_{0}$, $\{\theta_{k}\}$, $\{T_{k}^{\text{comp}}\}$, $T_{d}^{+} = \max\limits_{k}\{T_{k}^{\text{comp}}+R_{k}(B_{0}/K)\}$, $T_{d}^{-}=\max\limits_{k}\{T_{k}^{\text{comp}}\}$, accuracy threshold $\varepsilon$, and temporary variable $\bar{B}_{0}=0$.}
		\While{$\vert B_{0}-\bar{B}_{0} \vert > \varepsilon$}
		\State{$T_{d} = \frac{T_{d}^{+}+T_{d}^{-}}{2}$}
		\For{$k=1:K$}
		\State{Solve $R_{k}(b_{k})=\frac{S}{T_{d} - T_{k}^{\text{comp}}}$ by bisection search with respect to $b_{k}$.}
		\EndFor
		\State{$\bar{B}_{0} = \sum_{k=1}^{K}b_{k}$}
		\If{$\bar{B}_{0}>B_{0}$}
		\State{$T_{d}^{-}=T_{d}$}
		\Else
		\State{$T_{d}^{+} = T_{d}$}
		\EndIf 
		\EndWhile
		\Return{$\{b_{k}\}$}
	\end{algorithmic}
\end{algorithm}

\subsection{Quantization level optimization}\label{sec:optimization}

Next, we focus on optimizing the quantization level $q$ under fixed bandwidth allocation. First, we relax the value of $q$ from integer to real number in interval $q\in[2,+\infty)$. For the convenience of optimization, we approximate $N_{\epsilon}$ in Proposition \ref{coro} as $N_{\epsilon} \approx \frac{\sqrt{d}}{qK}H_{1} + H_{2}$, where $H_{1} = \frac{A}{\epsilon}-B$ and $H_{2} = \frac{A+D}{\epsilon}-B -C$, by getting rid of the ceiling operation. Then, we introduce an intermediate variable $\tilde{T}$, and Problem (P1) reduces to 
\begin{align}
	\label{eq:problem_quan}\text{(P3)}\;\;\;\underset{q,\tilde{T}}{\text{min   }}
	\;\;\;&\tilde{T},\\  \text{s.t.} \;\;\;
	&\nonumber \left(T_{k}^{\text{comp}}+T_{k}^{\text{comm}}\right)\left(\frac{\sqrt{d}}{qK}H_{1} + H_{2}\right)\leq \tilde{T},\tag{\ref{eq:problem_quan}a}\\
	&\nonumber q \ge 2.\tag{\ref{eq:problem_quan}b}
\end{align}

Problem (P3) is non-convex due to the non-convexity of constrains in (\ref{eq:problem_quan}a). To tackle this problem,
the successive convex approximation (SCA) technique can be applied to obtain a stationary point \cite{Razaviyayn2014SCA}. An algorithm summarizing the above procedure is given by \textbf{Algorithm 2}. The key idea is that in each iteration, the original problem is approximated by a tractable convex one at a given local point as elaborated below. To start with, we substitute $T_{k}^{\text{comm}}=\frac{\left(1+\log_{2}(1+q)\right)d}{R_{k}}$ into (\ref{eq:problem_quan}a), and obtain
\begin{equation*}
	\left(T_{k}^{\text{comp}}+\frac{\left(1+\log_{2}(1+q)\right)d}{R_{k}}\right)\left(\frac{\sqrt{d}}{qK}H_{1} + H_{2}\right)\leq \tilde{T}.
\end{equation*}
After taking the logarithm of both sides, and rearranging, it yields
\begin{equation}\label{eq:after_log}
	J_{k}(q) - \ln(qK) -\ln(\tilde{T}) \leq 0,
\end{equation}
where $J_{k}(q)=\ln\left(T_{k}^{\text{comp}}+\frac{\left(1+\log_{2}(1+q)\right)d}{R_{k}}\right) + \ln\left(qKH_{2}+H_{1}\sqrt{d}\right)$. It can be verified that $J_{k}(q)$ is a concave function of $q$. Recall that any concave function is globally upper-bounded by its first-order Taylor expansion at any point. Therefore, with given local point $q^{(r)}$, we can establish an upper bound of $J_{k}(q)$ as
\begin{equation*}
	J_{k}(q) \leq J_{k}(q^{(r)}) + J_{k}^{\prime}(q^{(r)})\left(q-q^{(r)}\right) \triangleq \hat{J}_{k}(q),
\end{equation*}
where $J_{k}^{\prime}(q^{(r)})$ is the derivative of $J_{k}(q)$ at $q^{(r)}$, i.e., 
\begin{align*}
	&J_{k}^{\prime}(q^{(r)})= \frac{KH_{2}}{q^{(r)}KH_{2}+H_{1}\sqrt{d}} \\
	&+ \frac{1}{\ln(2)\left(\log_{2}(1+q^{(r)})+R_{k}T_{k}^{\text{comm}}/d+1\right)(1+q^{(r)})}.
\end{align*}
By replacing $J_{k}(q)$ in (\ref{eq:after_log}) with its upper bound $\hat{J}_{k}(q)$, with given local point $q^{(r)}$ at $r$-th iteration, the next point at $(r+1)$-th iteration can be obtained by solving the following problem
\begin{align}
	&\label{eq:problem_quan_approx}\text{(P3.1)}\;\;\;q^{(r+1)}=\underset{q}{\arg\min}
	\;\;\;\tilde{T},\\  \text{s.t.} \;\;\;
	&\nonumber \hat{J}_{k}(q) - \ln(qK) -\ln(\tilde{T}) \leq 0,\;\forall k\in[K],\tag{\ref{eq:problem_quan_approx}a}\\
	&\nonumber q \ge 2.\tag{\ref{eq:problem_quan_approx}b}
\end{align}
Since the left side of constraint (\ref{eq:problem_quan_approx}a) are jointly convex with respect to $q$ and $\tilde{T}$, Problem (P3.1) is convex, which can be solved by standard convex optimization tools such as CVXPY \cite{cvxpy}. After the iterations converge, e.g., the gap between $\tilde{T}^{(r)}$ and $\tilde{T}^{(r+1)}$ is lower than a given threshold, Problem (P3) is deemed solved.

\begin{algorithm}\small
	\caption{SCA method for solving Problem (P3)}
	\label{algo:SCA}
	\begin{algorithmic}[1]
		\State {Find a feasible initial quantization level $q^{(0)}$ in (\ref{eq:problem_quan}), and set $r=0$ and the threshold $\varepsilon$.}
		\Repeat
		\State {Set $q^{(r+1)}$ as the solution of Problem (P3.1).}
		\State {$r$ $\leftarrow$ $r+1$}
		\Until{$\vert \tilde{T}^{(r)}-\tilde{T}^{(r-1)} \vert \leq \varepsilon$}
		\Return $q^{(r)}$
	\end{algorithmic}
\end{algorithm}

\subsection{A joint optimization algorithm}

Since the unique solution of Problem (P2) can be obtained by \textbf{Algorithm 1} and a stationary point of Problem (P3) can be reached by \textbf{Algorithm 2}, the sub-optimal bandwidth allocation and quantization level can be jointly obtained by alternately solving Problem (P2) and (P3), which is summarized in \textbf{Algorithm 3}. It can be verified that the sub-optimal solution from \textbf{Algorithm 3} is a stationary point of the original Problem (P1).

Note that solving Problem (P1) may lead to non-integer $q$, which needs further rounding technique to yield an integer $q$ for practical implementation. One possible rounding technique is discussed as follows. We denote $T(q, \{b_{k}\})$ as the total training time in Problem (P1) when $q$ and $\{b_{k}\}$ is substituted. After finding the optimized quantization level $\hat{q}$ and optimal bandwidth allocation $\{b_{k}^{\ast}\}$, the final quantization level $q^{\ast}$ is obtained as
\begin{equation*}
	q^{\ast}=\arg\min_{q\in\{\lceil \hat{q}\rceil-1,\lceil \hat{q}\rceil\}}T(q, \{b_{k}^{\ast}\}).
\end{equation*}

\begin{algorithm}\small
	\caption{The joint optimization algorithm for solving Problem (P1)}
	\label{algo:joint}
	\begin{algorithmic}[1]
		\State{Initialization: Quantization level $q^{(0)}$, bandwidth allocation $\{b_{k}^{(0)}\}$. Set $r=0$ and the accuracy threshold $\varepsilon$.}
		\Repeat
		\State{Update the bandwidth allocation$\{b_{k}^{r+1}\}$ by \textbf{Algorithm 1}.}
		\State{Update the quantization level $q^{(r+1)}$ and total training time $\tilde{T}(q^{(r+1)}, \{b_{k}^{(r+1)}\})$ by \textbf{Algorithm 2}.}
		\State{$r$ $\leftarrow$ $r+1$}
		\Until{$\vert T(q^{(r)}, \{b_{k}^{(r)}\}) - T(q^{(r-1)}, \{b_{k}^{(r-1)}\}) \vert \leq \varepsilon$}
		\Return{$\{b_{k}^{\ast} = b_{k}^{(r)}\}$ and $q^{\ast}=\arg\min_{q\in\{\lceil q^{(r)}\rceil-1,\lceil q^{(r)}\rceil\}}T(q, \{b_{k}^{\ast}\})$}
	\end{algorithmic}
\end{algorithm}

\section{Experimental evaluation}\label{sec:experiment}


In this section, we provide numerical results of two experiments under different wireless communication scenarios and learning tasks, which capture real system heterogeneity, to examine our theoretical results. In Experiment I, we consider a learning task with strongly convex loss function and training model of small size. In Experiment II, to stretch the theory, we consider a learning task with non-convex loss function and training model of large size. Although our analysis is developed based on the assumption of strongly convex loss function, we show that the proposed algorithm can also work well in the case with non-convex loss function. All experiments are implemented by PyTorch using Python 3.8 on a Linux server with one NVIDIA\textsuperscript{\textregistered} GeForce\textsuperscript{\textregistered} RTX 3090 GPU 24GB and one Intel\textsuperscript{\textregistered} Xeon\textsuperscript{\textregistered} Gold 5218 CPU.

\subsection{Experiment setup}

\textbf{FEEL system}: We consider a FEEL system with a edge server covering a circular area of radius $r=500$ m. Within the area, $K=6$ edge devices are placed randomly and distributed uniformly over the circular area with the exclusion of a central disk of radius $r_{h}=100$ m. The transmit power of each device is $1$ dBm. To expose the heterogeneity of the edge devices, the CPU frequency of each device is assumed to be uniformly distributed from $100$ MHz to 1 GHz. The number of processing cycles of device $k$ for executing one batch of samples is $\nu = 10^{8}$ in Experiment I and $\nu = 2.5 \times 10^{10}$ in Experiment II.

\textbf{Wireless propagation}: The large-scale propagation coefficient in dB from device $k$ to the edge server is modeled as $[\phi_{k}]_{\text{dB}}=[\text{PL}_{k}]_{\text{dB}} +  [\zeta_{k}]_{\text{dB}}$, where $[\text{PL}_{k}]_{\text{dB}} = 128.1+37.6\log_{10}\text{dist}_{k}$ ($\text{dist}_{k}$ is the distance in kilometer) is the path loss in dB, and $[\zeta_{k}]_{\text{dB}}$ is the shadow fading in dB \cite{Yang2020TWC-pathloss}. In this simulation, $[\zeta_{k}]_{\text{dB}}$ is Gauss-distributed random variable with mean zero and variance $\sigma^{2}_{\zeta} = 8$ dB. The noise power spectral density is $N_{0}=-174$ dBm/Hz, and the total bandwidth is $B_{0}=10$ KHz \cite{Dhillon2017Bandwidth}.

\textbf{Learning tasks and models}: In Experiment I, we consider the $\ell_{2}$ regularized logistic regression task on synthetic data \cite{Wang2018NIPS_spa}. The local loss function in (\ref{eq:local_loss_function}) at device $k$ is given by
\begin{equation*}
	F_{k}(\mathbf{w})= \frac{1}{D}\sum_{(\mathbf{x}_{i},y_{i})\in\mathcal{D}_{k}}\log_{2}\left(1 + \exp(-\mathbf{x}_{i}^{T}\mathbf{w}y_{i})\right) + \lambda \Vert \mathbf{w} \Vert_{2}^{2},
\end{equation*}
where $\mathbf{x}_{i} \in \mathbb{R}^{d}$ and $y_{i} \in \{-1,1\}$. The $\ell_{2}$ regularization parameter $\lambda$ is set to $\lambda=10^{-6}$. It can be verified that the local loss function $F_{k}(\mathbf{w})$ is smooth and strongly convex. Each data sample $(\mathbf{x}_{i},y_{i})$ is generated in four steps as follows
\begin{enumerate}
	\item Dense data generation: $\bar{x}_{ij} \sim \mathcal{N}(0,1)$, $\forall\;j\in[d]$;
	\item Magnitude sparsification: $\Theta_{j} \sim \text{Uniform}[0,1]$, $\Theta_{j} \leftarrow \Delta_{1} \Theta_{j}$ if $\Theta_{j} \leq \Delta_{2}$, $\forall\;j\in[d]$;
	\item Data sparsification: $x_{ij} \leftarrow \bar{x}_{ij} \cdot \Theta_{j}$, $\forall\;j\in[d]$;
	\item Label generation: $\mathbf{w} \sim \mathcal{N}(0,\mathbf{I}_{d})$, $y_{i} \leftarrow \mathrm{sgn}(\bar{\mathbf{x}}_{i}^{T}\mathbf{w})$.
\end{enumerate}
Note that the parameters $\Delta_{1}$ and $\Delta_{2}$ control the sparsity of data points and the gradients\footnote{From our experiments, the effect of stochastic quantization on SGD convergence depends heavily on the sparsity structure of the gradients. Therefore, we choose this dataset in Experiment I to better validate our theoretical results. Moreover, to the best of our knowledge, how the sparsity structure of the gradients affects the learning algorithm that employs stochastic quantization as compression scheme has not been revealed in the literature, which is an interesting topic but beyond the scope of this work.} \cite{Wang2018NIPS_spa}. The parameters $\Delta_{1}$ and $\Delta_{2}$ are set to $0.9$ and $0.25$ in this experiment. Moreover, the dimension of each data point is set to $d=1024$. Hence, the model contains 1024 parameters in total. We generate 48,000 data points for training and 12,000 data points for validation. 

In Experiment II, we consider the learning task of image classification using the well-known CIFAR-10 dataset, which consists of 50,000 training images and 10,000 validation images in 10 categories of colorful objectives such as airplanes, cars, etc. ResNet-20 (269,722 parameters in total) with batch normalization\footnote{The implementation of ResNet-20 follows this GitHub repository: \url{https://github.com/hclhkbu/GaussianK-SGD} \cite{Shi2019ResNet20}.} is applied as the classifier model \cite{He2016ResNet}.

\textbf{Training and optimization parameters}: We consider a decaying learning rate as $\eta_{n} = \frac{5}{n+10}$ in Experiment I, where $n$ is the index of communication round, and the learning rate is set to $\eta_{n} = \frac{100}{n+1000}$ in Experiment II. To deliver rigorous results, we strictly control all unrelated variables in both experiments.

\subsection{Experiment results in Experiment I}


\subsubsection{Estimation of data-related parameters} 
\begin{figure}[!t]
	\centering
	\includegraphics[width=0.48\textwidth]{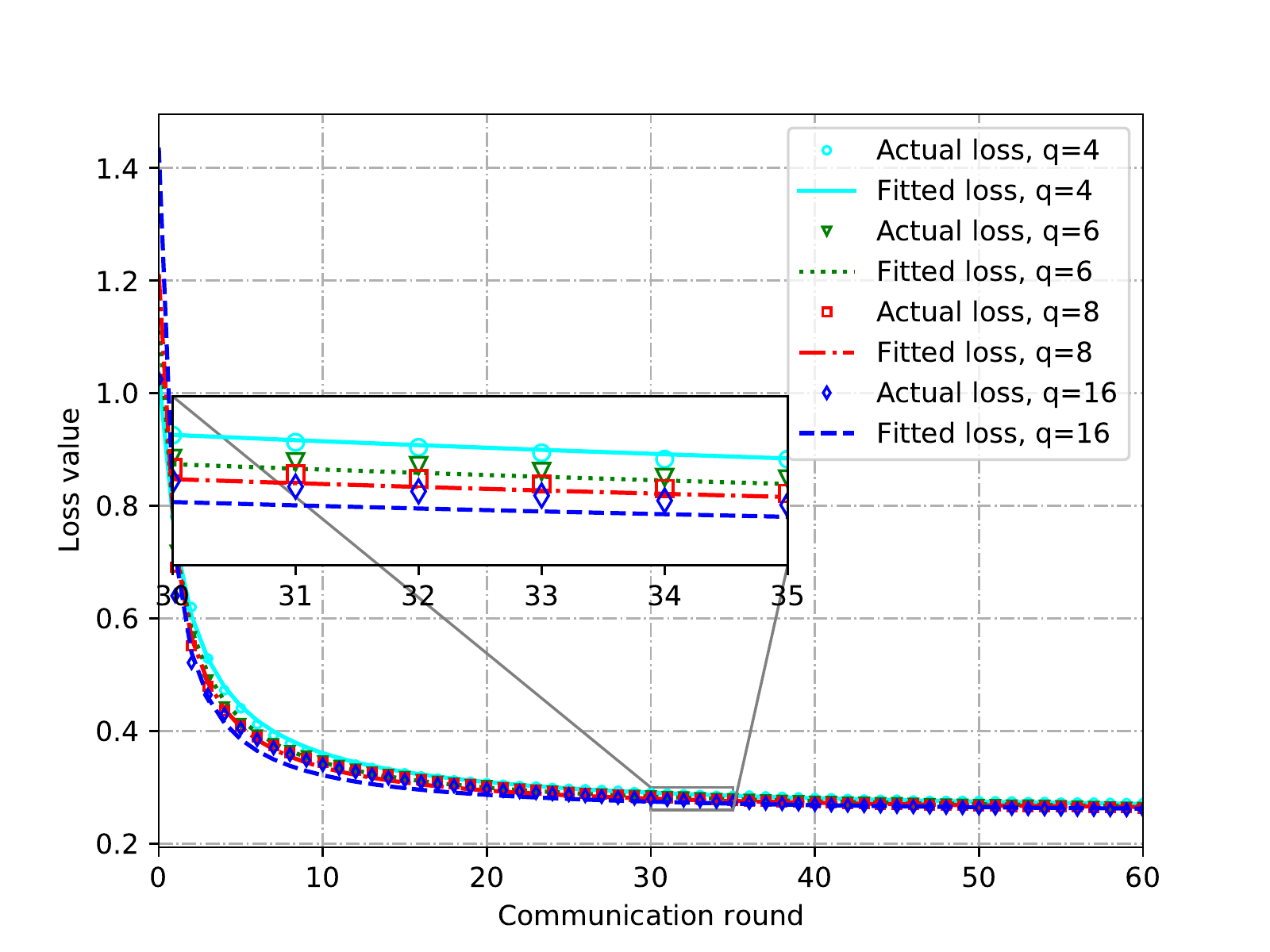}
	\caption{Robustness of the joint-data-and-model-driven fitting based method. (The fitted loss function and the actual loss versus communication round when the quantization levels are set as $q_1$ and $q_2$ and other values.)}
	\label{fig:re_fig_fitting_sd}
\end{figure}

In the optimization in Section \ref{sec:optimization}, we need to obtain the values of $H_1$ and $H_2$ using the proposed joint data-and-model-driven fitting method in Section \ref{sec:commun-rounds}. With the joint data-and-model-driven fitting method, for any given two quantization levels $q_1$ and $q_2$, and the threshold of loss optimality gap upper bound $\epsilon$, we can obtain an estimation of $H_1$ and $H_2$, which are used in \textbf{Algorithm 3}. Also, the optimal loss value can be obtained by the estimation of $Z$ in Eq. (\ref{eq:estimation_upper_bound}), i.e., $F(\mathbf{w}_{\ast}) \approx Z \approx 0.247$.
The threshold of loss optimality gap upper bound is set as $\epsilon = 0.012$. 
One can choose any combination of $q_1$ and $q_2$ to implement the estimation in theory. As a reminder to the readers, however, in our experience, the combination of $q_1$ and $q_2$ with large difference leads to better estimation accuracy.
In our results, we choose $(q_1, q_2) = (4,6)$ in the joint data-and-model-driven fitting method, and obtain that $H_1 \approx 43.01$ and $H_2 \approx 48.79$. To show the robustness of our estimation method, we plot the fitted loss function and the actual loss when the quantization levels are $q_1 = 4$ and $q_2 = 6$, and also other values than $q_1$ and $q_2$, e.g., $8$ and $16$, as shown in Fig. \ref{fig:re_fig_fitting_sd}, and we can see that the fitted loss function fits the actual loss well.

\subsubsection{Optimization of quantization level} 
\begin{figure}[!t]
	\centering
	\includegraphics[width=0.48\textwidth]{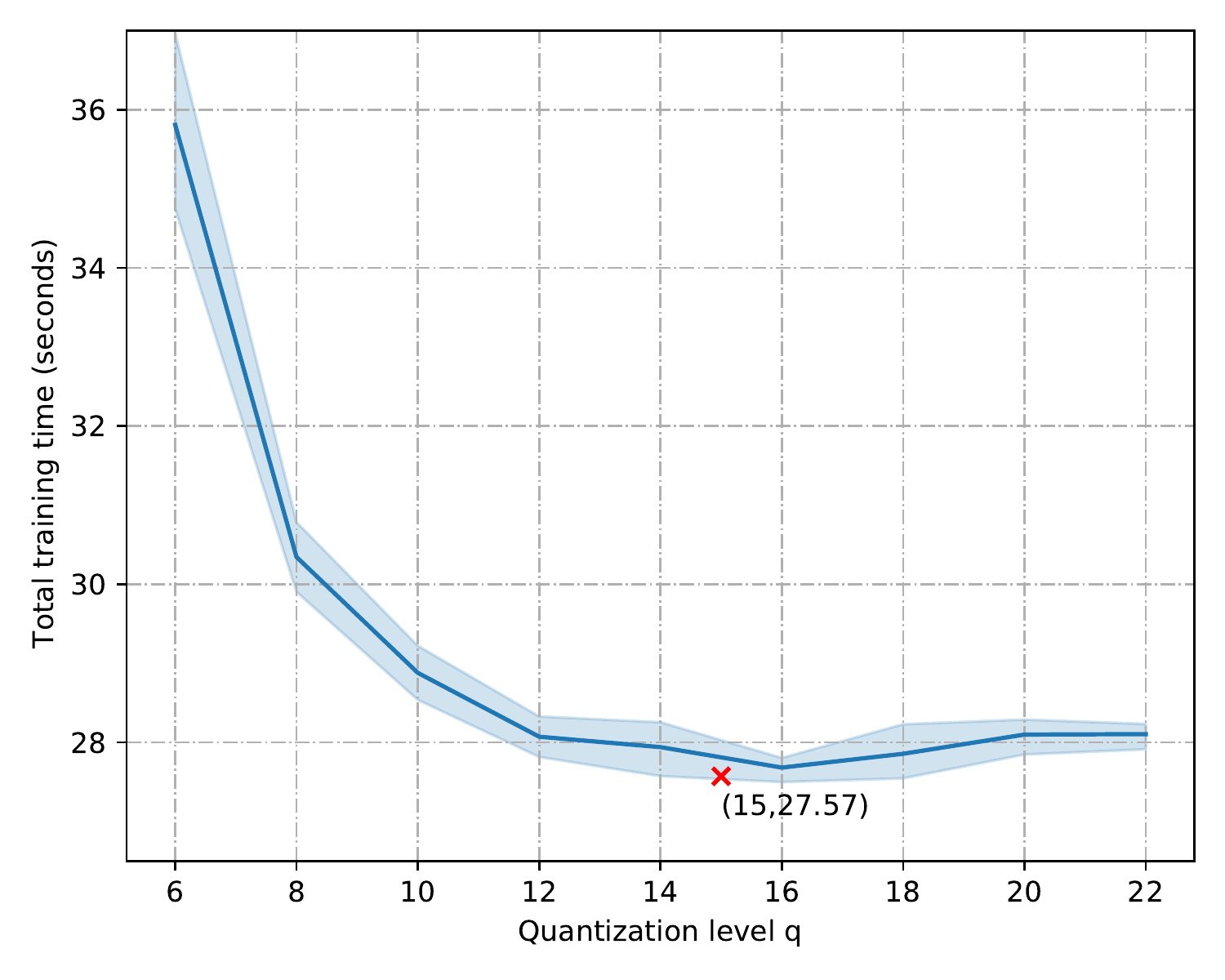}
	\caption{The total training time versus quantization level $q$ in simulation when the bandwidth allocation is optimal in Experiment I. (The optimal quantization level and corresponding training time from \textbf{Algorithm 3} in theory are annotated by ``x'' in red.)}
	\label{fig:fig_quan_opt_mnist_iid}
\end{figure}

\begin{figure}[!htb]\small
	\centering
	\begin{tabular}{c}
		\includegraphics[width=0.48\textwidth]{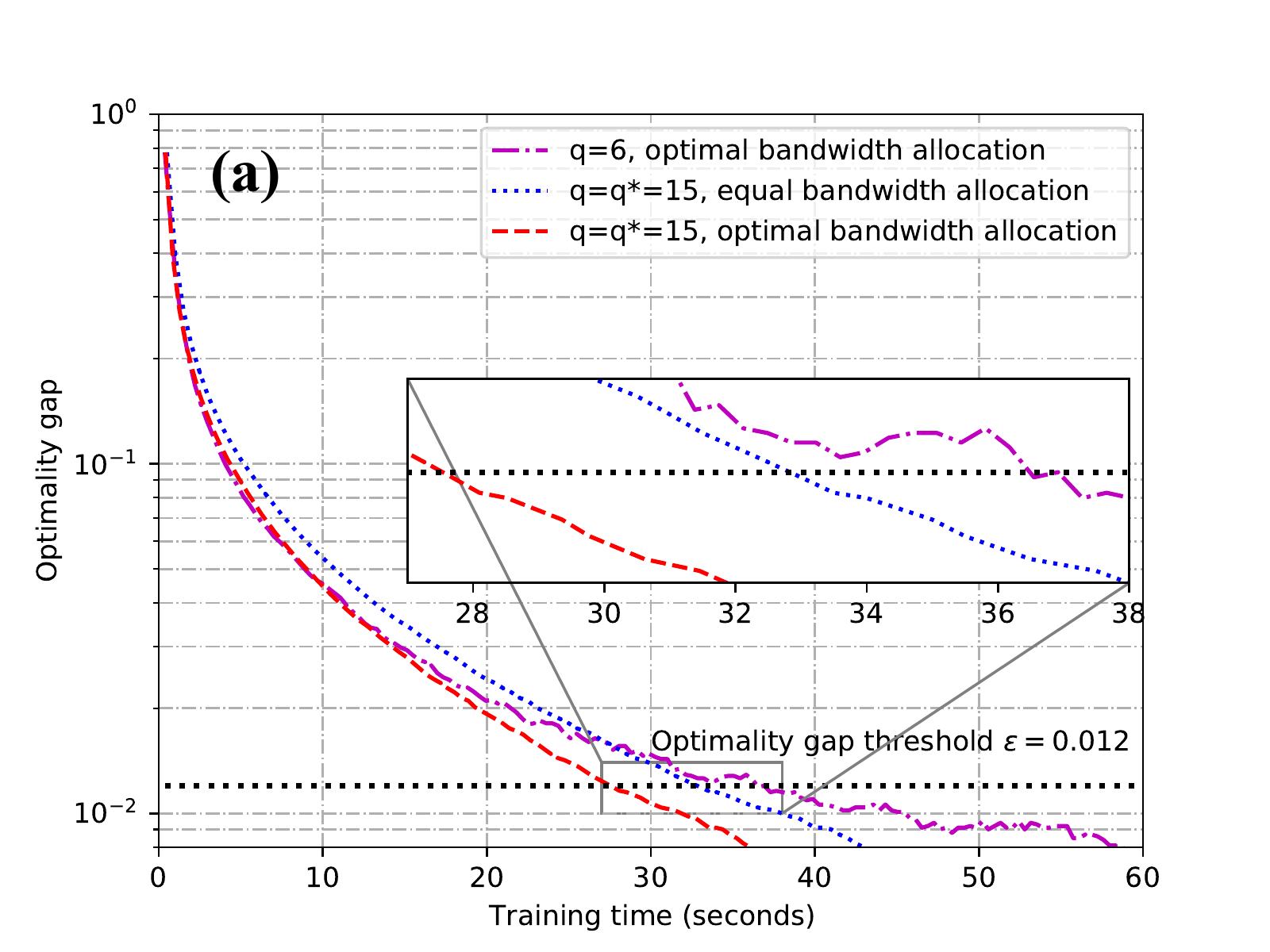}\\
		\includegraphics[width=0.48\textwidth]{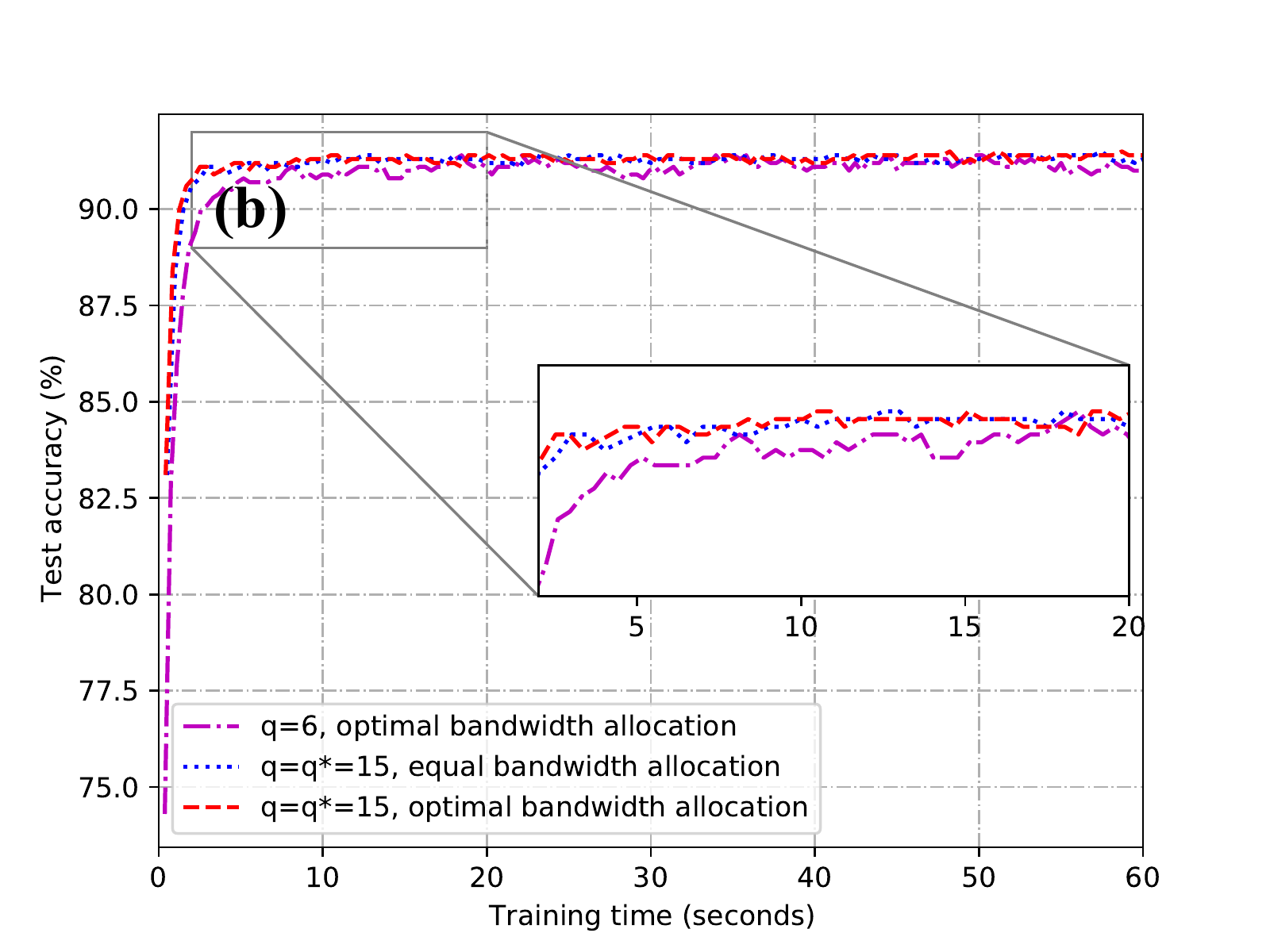}\\
	\end{tabular}
	\caption{Optimality gap versus training time (a) and test accuracy versus training time (b)  in Experiment I}
	\label{fig_loss_acc}
\end{figure}

Fig. \ref{fig:fig_quan_opt_mnist_iid} plots the total training time versus communication round in simulation when the bandwidth allocation is optimal. We run the same training process for at least 5 times on each quantization level.
Fig. \ref{fig:fig_quan_opt_mnist_iid} shows that there exists optimal quantization level that minimizes the total training time. Recall the facts from Section \ref{sec:train_time_analysis} that the total training time $T=N_{\epsilon} \cdot T_d$, and $N_{\epsilon}$ is an decreasing function of quantization level $q$, while $T_d$ is an increasing function of $q$.
In other words, Fig. \ref{fig:fig_quan_opt_mnist_iid} demonstrates the trade-off between total communication rounds $N_{\epsilon}$ and per-round latency $T_d$ in FEEL system.
Moreover, it can be observed from the figure that the optimal quantization levels obtained in theory from \textbf{Algorithm 3} in Section \ref{sec:time_optimization} match the results by simulation, which confirms the validity of our proposed algorithms.
In Fig. \ref{fig_loss_acc}(a), we compare the training loss under optimal quantization level and optimal bandwidth allocation and the training loss under other quantization level and optimal bandwidth allocation. It can be observed that the training loss of optimal quantization level under optimal bandwidth allocation reaches the predefined threshold in a shorter time. 

\subsubsection{Optimization of bandwidth} 
\begin{figure}[!t]
	\centering
	\includegraphics[width=0.45\textwidth]{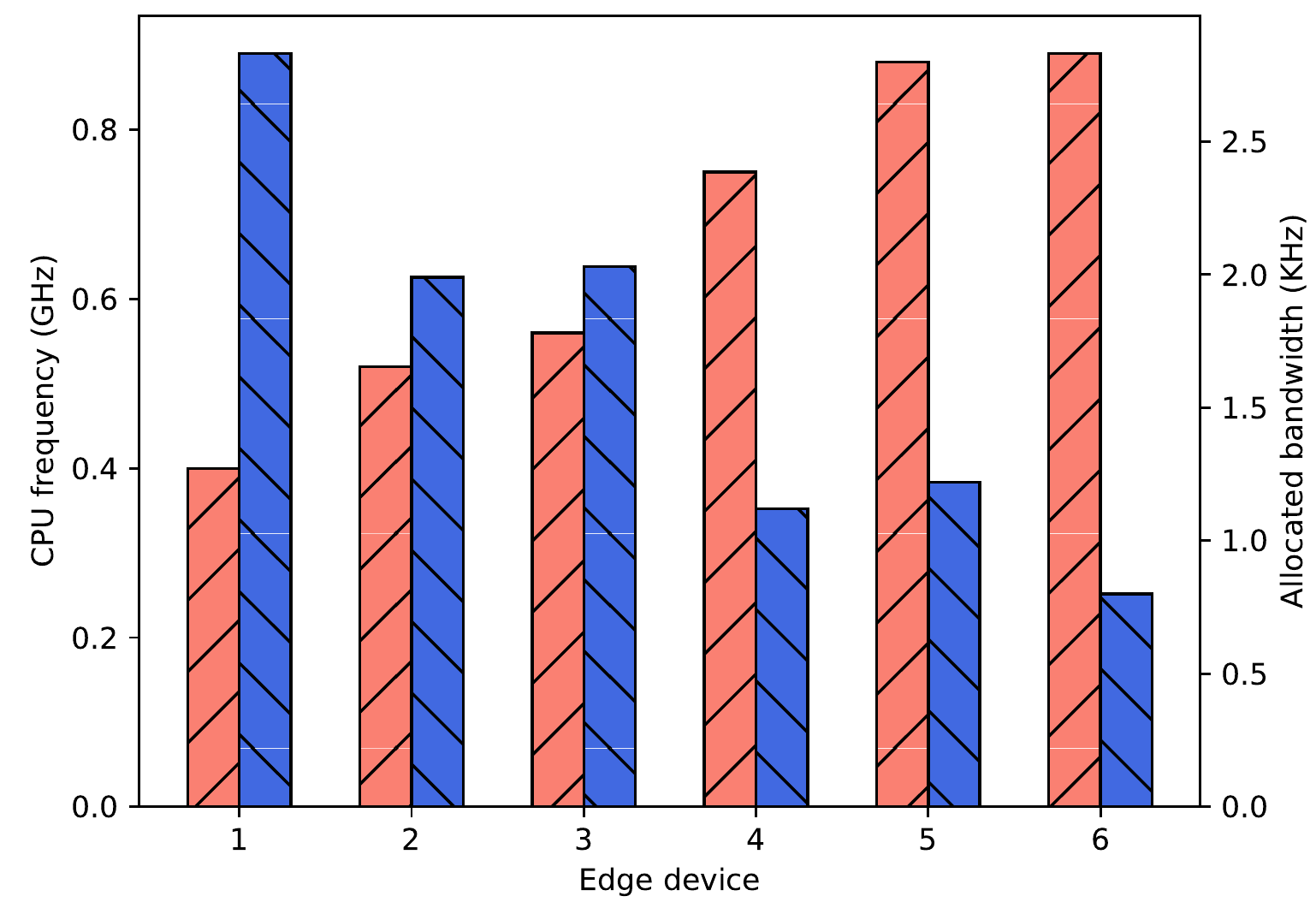}
	\caption{Optimal bandwidth allocation (bars on the right) and CPU frequency (bars on the left) of each edge device in Experiment I.}
	\label{fig:fig_bar_sd}
\end{figure}

Fig. \ref{fig_loss_acc} depicts the comparison between the schemes with optimal and equal bandwidth allocation in terms of loss optimality gap and test accuracy. We can observe that the scheme with optimal bandwidth allocation can reach the predefined threshold and obtain a higher test accuracy in a shorter time. This indicates that our bandwidth allocation algorithm is effective and necessary in FEEL system. To show how the heterogeneous computation power of edge devices affect the communication resource allocation, we present the CPU frequency of each edge device and its corresponding optimal allocated bandwidth in Fig. \ref{fig:fig_bar_sd}. It can be observed that the edge devices with lower CPU frequency will be allocated with a larger bandwidth, which in spirit has similarity to the well-known phenomenon of ``water-filling'' in the problem of power allocation in wireless communication \cite{Gong2010TSP}. 

\subsection{Experiment results in Experiment II}

\begin{figure}[!t]
	\centering
	\includegraphics[width=0.48\textwidth]{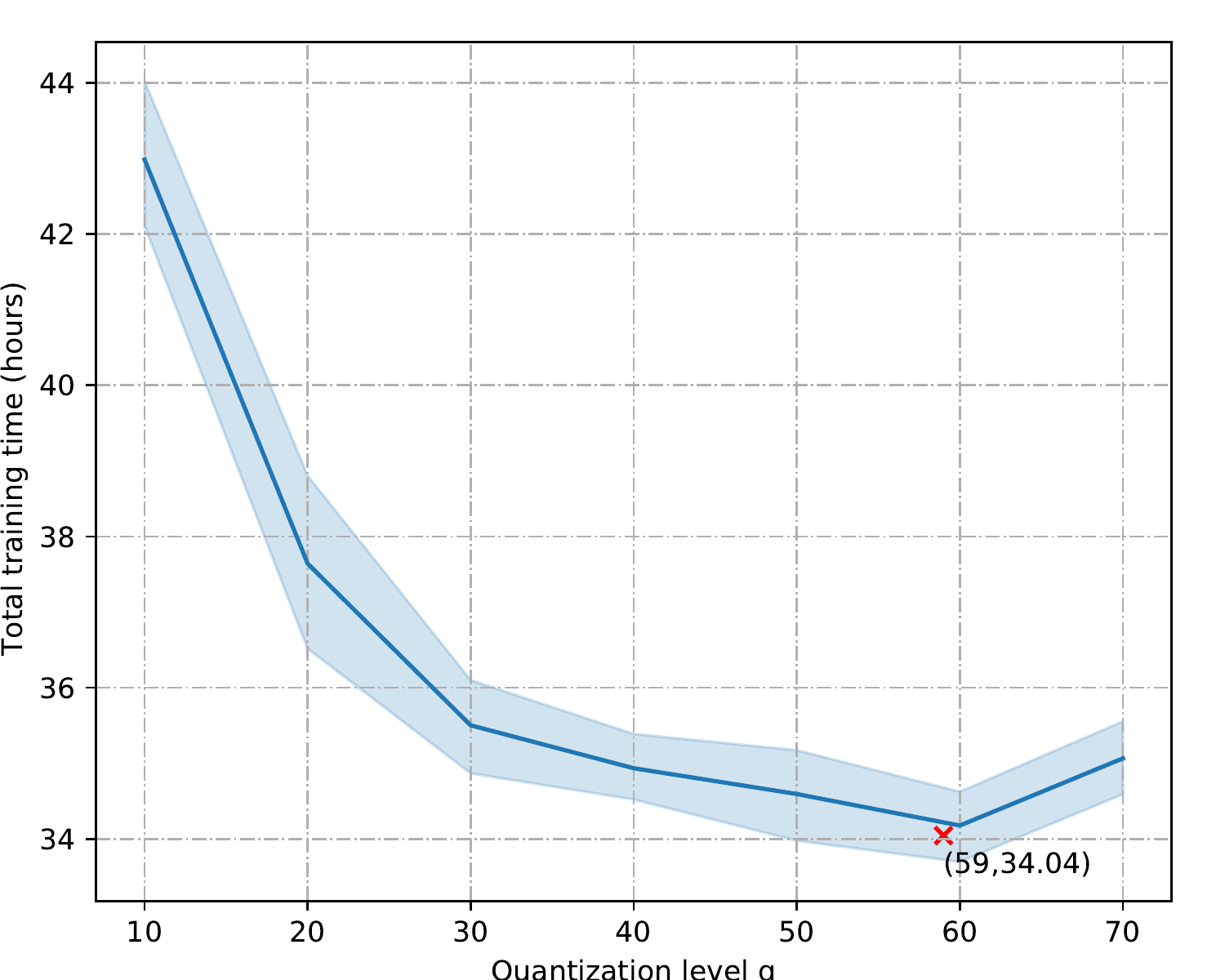}
	\caption{The total training time versus quantization level $q$ in simulation when the bandwidth allocation is optimal in Experiment II. (The optimal quantization levels and corresponding training time from \textbf{Algorithm 3} in theory are annotated by ``x'' in red.)}
	\label{fig:re_fig_quan_opt_resnet}
\end{figure}

\begin{figure}[!htb]\small
	\centering
	\begin{tabular}{c}
		\includegraphics[width=0.48\textwidth]{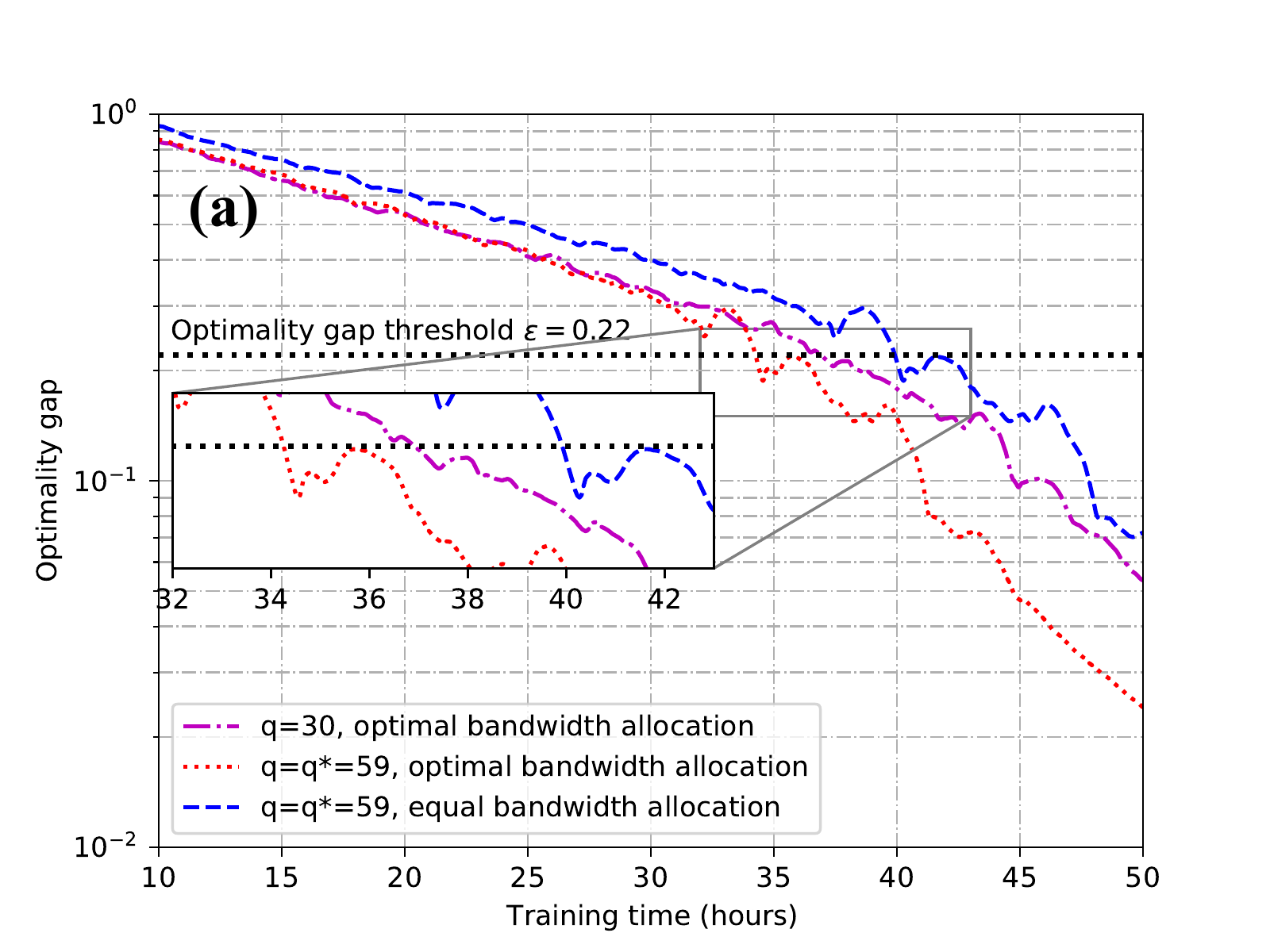}\\
		\includegraphics[width=0.48\textwidth]{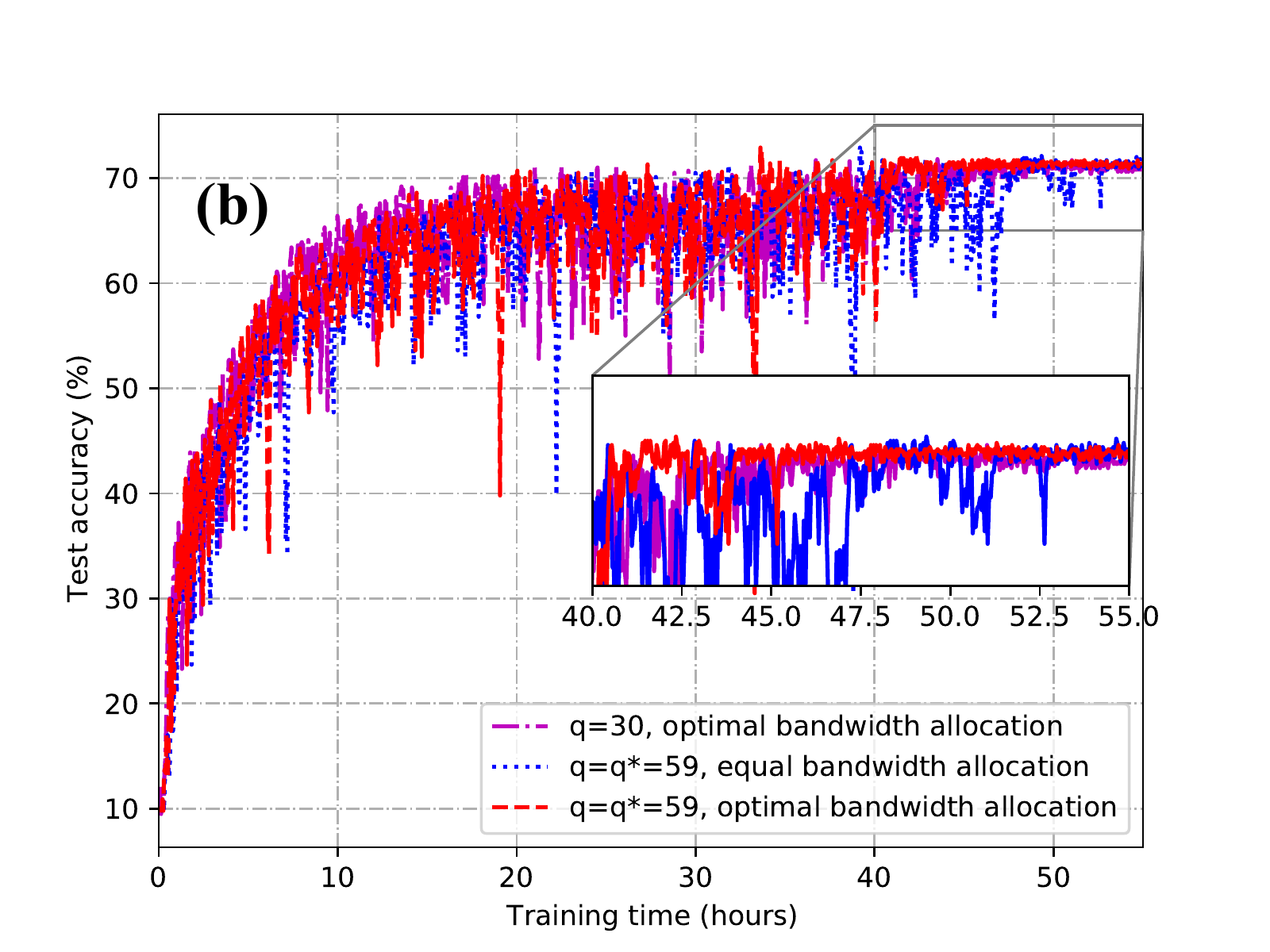}\\
	\end{tabular}
	\caption{Optimality gap versus training time (a) and test accuracy versus training time (b)  in Experiment II}
	\label{fig_gap_acc_cnn}
\end{figure}

We conduct Experiment II to evaluate our method and algorithms on learning model with non-convex loss function. In this experiment, we choose $(q_1, q_2) = (15,20)$ and obtain $H_1 \approx 96.26$ and $H_2 \approx 808.53$ by our joint data-and-model-driven fitting method. The threshold of loss optimality gap upper bound is set as $\epsilon = 0.22$. 
Fig. \ref{fig:re_fig_quan_opt_resnet} shows the total training time versus quantization level in simulation when the bandwidth allocation is optimal; Fig. \ref{fig_gap_acc_cnn} shows the loss value and test accuracy versus training time in simulation with different quantization levels and bandwidth allocations. We can obtain the similar observations from Fig. \ref{fig:re_fig_quan_opt_resnet} and Fig. \ref{fig_gap_acc_cnn} as in Experiment I, which reveals that our method and algorithm also work well in non-convex setting, although they are derived under strongly-convex setting. 

\section{Conclusion}\label{sec:conclusion}

This paper have studied the training time minimization for quantized FEEL with optimized quantization level and bandwidth allocation.
On the basis of the convergence analysis of quantized FEEL and our proposed joint-data-and-model-driven fitting method, we derived the closed-form expression of the total training time and characterizes the trade-off between the convergence speed and communication overhead, which is governed by the quantization level.
Next, we minimized the total training time by optimizing the quantization level and bandwidth allocation, for which a high-quality near-optimal solutions are obtained by alternating optimization.
The theoretical results developed can be used to guide the system optimization and contribute to the understanding of how wireless communication system can properly coordinate resources to accomplish learning tasks. 
This also opens several directions for further research. One direction is to implement device sampling in quantized FEEL, in which how the bandwidth is allocated to minimize the training time is completely a different story. The other direction is to consider error compensation in quantized FEEL to mitigate the effects of compression errors.

\appendices

\section{Proof of Theorem \ref{theorem:convergence}}\label{apdx:theorem_1}

We first give one necessary lemma before proving Theorem \ref{theorem:convergence}, and also provide the proof for this lemma in Appendix \ref{apdx:lemma_1}.

\begin{lemma}\label{lemma:convergence_eta_2}
	Under the same conditions in Theorem \ref{theorem:convergence}, we have
	\begin{align*}
		\mathbb{E}\left[\left\Vert \mathbf{w}^{(n+1)} - \mathbf{w}_{\ast} \right\Vert^{2}\right] &\leq \left(1 - \mu\eta_{n}\right)\mathbb{E}\left[\left\Vert \mathbf{w}^{(n)} - \mathbf{w}_{\ast} \right\Vert^2\right] \\&+ \alpha\Gamma\eta_{n}^{2},
	\end{align*}
	where $\Gamma = 2 LF_{\delta} +\frac{1}{K}\sum_{k=1}^{K}\delta_{k}^{2}$.
\end{lemma}

Next, with Lemma \ref{lemma:convergence_eta_2} and decaying learning rate $\eta_{n} = \frac{\beta}{\gamma + n}$, we prove that $\mathbb{E}\left[\left\Vert \mathbf{w}^{(n)} - \mathbf{w}_{\ast} \right\Vert^2\right] \leq \frac{\nu}{\gamma + n}$ by induction where
\begin{equation*}
	\nu = \max\left\lbrace(\gamma+1)\left\Vert \mathbf{w}_{0} - \mathbf{w}_{\ast} \right\Vert^2,\frac{\alpha \Gamma\beta^{2}}{\mu\beta-1}\right\rbrace.
\end{equation*} 
First, it holds for $n=1$ by the definition of $\nu$. Then, assuming it holds for some $n>1$, it follows from Lemma \ref{lemma:convergence_eta_2} that
\begin{align*}
	&\mathbb{E}\left[\left\Vert \mathbf{w}^{(n+1)} - \mathbf{w}_{\ast} \right\Vert^{2}\right]\\
	\leq & \frac{(\gamma+n-\mu\beta)\nu}{(\gamma+n)^{2}} + \frac{\alpha \Gamma\beta^{2}}{(\gamma+n)^{2}}\\
	= & \frac{(\gamma+n-1)\nu}{(\gamma+n)^{2}} + \frac{\alpha \Gamma\beta^{2}-(\mu\beta-1)\nu}{(\gamma+n)^{2}}.
\end{align*}
By the definition of $\nu$, we have $\alpha \Gamma\beta^{2}-(\mu\beta-1)\nu \leq
0$. Then, it follows that
\begin{align*}
	\mathbb{E}\left[\left\Vert \mathbf{w}^{(n+1)} - \mathbf{w}_{\ast} \right\Vert^{2}\right] &\leq 
	\frac{(\gamma+n-1)\nu}{(\gamma+n)^{2}}\\
	&\leq \frac{\nu}{\gamma+n+1}.
\end{align*}
Specifically, we choose $\beta = \frac{2}{\mu}$, $\gamma = \frac{2\alpha L}{\mu}-1$. Using $\max\{x,y\}\leq x+y$, we have $\nu \leq \frac{2\alpha L}{\mu}\left\Vert \mathbf{w}_{0} - \mathbf{w}_{\ast} \right\Vert^2 + \frac{4\alpha \Gamma}{\mu^{2}}$. Therefore, it satisfies that
\begin{align*}
	&\mathbb{E}\left[\left\Vert \mathbf{w}^{(n)} - \mathbf{w}_{\ast} \right\Vert^{2}\right] \\
	\leq & \frac{\alpha/\mu}{n+2\alpha L/\mu-1}\left(2L\left\Vert \mathbf{w}_{0} - \mathbf{w}_{\ast} \right\Vert^2 + \frac{4\Gamma}{\mu}\right).
\end{align*}
Then, by the $L$-smoothness of $F(\mathbf{w})$, it holds that
\begin{equation*}
	\mathbb{E}\left[F(\mathbf{w}^{(n)})\right] - F(\mathbf{w}_{\ast}) \leq \frac{L}{2}\mathbb{E}\left[\left\Vert \mathbf{w}^{(n)} - \mathbf{w}_{\ast} \right\Vert^{2}\right].
\end{equation*}
It follows that 
\begin{align*}
	& \mathbb{E}\left[F(\mathbf{w}^{(n)})\right] - F(\mathbf{w}_{\ast}) \\
	\leq & \frac{\alpha L/\mu}{n+2\alpha L/\mu-1}\left(L\left\Vert \mathbf{w}_{0} - \mathbf{w}_{\ast} \right\Vert^2 + \frac{2\Gamma}{\mu}\right).
\end{align*}
We complete the proof of Theorem \ref{theorem:convergence} by setting $n=N$.


\section{Proof of Lemma \ref{lemma:convergence_eta_2}}\label{apdx:lemma_1}
Notice that $\mathbf{w}^{(n+1)} = \mathbf{w}^{(n)} - \frac{\eta_{n}}{K}\sum_{k=1}^{K}\mathcal{Q}(\mathbf{g}_{k}^{(n)})$, then
\begin{align*}
	\nonumber & \left\Vert \mathbf{w}^{(n+1)} - \mathbf{w}_{\ast} \right\Vert^{2}\\
	=&\left\Vert \mathbf{w}^{(n)} - \frac{\eta_{n}}{K}\sum_{k=1}^{K}\mathcal{Q}(\mathbf{g}_{k}^{(n)}) - \mathbf{w}_{\ast} \right\Vert^{2}\\
	=&\left\Vert \mathbf{a}_{1} - \mathbf{a}_{2}\right\Vert^{2}\\
	=& \left\Vert \mathbf{a}_{1}\right\Vert^{2} + \left\Vert \mathbf{a}_{2}\right\Vert^{2} - 2\left\langle \mathbf{a}_{1}, \mathbf{a}_{2} \right\rangle,
\end{align*}
where $\mathbf{a}_{1} = \mathbf{w}^{(n)} - \mathbf{w}_{\ast} - \frac{\eta_{n}}{K}\sum_{k=1}^{K}\mathbf{g}_{k}^{(n)}$, and $\mathbf{a}_{2} = \frac{\eta_{n}}{K}\sum_{k=1}^{K}\left(\mathcal{Q}(\mathbf{g}_{k}^{(n)}) - \mathbf{g}_{k}^{(n)}\right)$. 
Due to $\mathbb{E}_{\mathcal{Q}}[\mathcal{Q}(\mathbf{g}_{k}^{(n)})]=\mathbf{g}_{k}^{(n)}$, we have $\mathbb{E}_{\mathcal{Q}}[\left\langle \mathbf{a}_{1}, \mathbf{a}_{2} \right\rangle]=0$, which leads to
\begin{equation}\label{eq:w_t_plus_1}
	\left\Vert \mathbf{w}^{(n+1)} - \mathbf{w}_{\ast} \right\Vert^{2} = \left\Vert \mathbf{a}_{1}\right\Vert^{2} + \left\Vert \mathbf{a}_{2}\right\Vert^{2}.
\end{equation}
Next, we first obtain upper bounds of $A_1$ and $A_2$; taking these bounds into (\ref{eq:w_t_plus_1}), then we find the connection between $\left\Vert \mathbf{w}^{(n+1)} - \mathbf{w}_{\ast} \right\Vert^{2}$ and $\left\Vert \mathbf{w}^{(n)} - \mathbf{w}_{\ast} \right\Vert^{2}$ after some proper manipulations.

\subsection{\textbf{Bound of} $\left\Vert \mathbf{a}_{1}\right\Vert^{2}$}To bound $\left\Vert \mathbf{a}_{1}\right\Vert^{2}$, we break $\left\Vert \mathbf{a}_{1}\right\Vert^{2}$ as
\begin{align*}
	\left\Vert \mathbf{a}_{1}\right\Vert^{2} &= \left\Vert  \mathbf{w}^{(n)} - \mathbf{w}_{\ast} - \frac{\eta_{n}}{K}\sum_{k=1}^{K}\mathbf{g}_{k}^{(n)}\right\Vert^{2}\\
	&= \left\Vert \mathbf{w}^{(n)} - \mathbf{w}_{\ast} \right\Vert^2 + \underbrace{\left\Vert \frac{\eta_{n}}{K}\sum_{k=1}^{K}\mathbf{g}_{k}^{(n)} \right\Vert^2}_{B_{1}} \\
	&+ \underbrace{2\left\langle \mathbf{w}_{\ast}-\mathbf{w}^{(n)},\frac{\eta_{n}}{K}\sum_{k=1}^{K}\mathbf{g}_{k}^{(n)} \right\rangle}_{B_{2}}.
\end{align*}
To bound $B_1$, we use $\Vert\sum_{k=1}^{K}\mathbf{a}_{k}\Vert^{2} \leq K\sum_{k=1}^{K}\Vert\mathbf{a}_{k}\Vert^{2}$. This gives
\begin{align*}
	B_1 \leq \frac{\eta_{n}^{2}}{K}\sum_{k=1}^{K}\Vert\mathbf{g}_{k}^{(n)}\Vert^{2}.
\end{align*}
By the $\mu$-strongly convexity of $F_{k}(\mathbf{w})$, it follows that
\begin{align*}
	&\left\langle \mathbf{w}_{\ast}-\mathbf{w}^{(n)},\mathbf{g}_{k}^{(n)} \right\rangle \\
	\leq& F_{k}(\mathbf{w}_{\ast}) - F_{k}(\mathbf{w}^{(n)}) - \frac{\mu}{2}\left\Vert \mathbf{w}^{(n)} - \mathbf{w}_{\ast} \right\Vert^2.
\end{align*}
Hence, $B_2$ can be bounded by
\begin{align*}
	B_{2} &\leq 2\frac{\eta_{n}}{K}\sum_{k=1}^{K}\left(F_{k}(\mathbf{w}_{\ast}) - F_{k}(\mathbf{w}^{(n)})\right) \\
	&- \mu\eta_{n}\left\Vert \mathbf{w}^{(n)} - \mathbf{w}_{\ast} \right\Vert^2.
\end{align*}

\subsection{\textbf{Bound of} $\left\Vert\mathbf{a}_{2}\right\Vert_{2}$} Since $\mathbf{g}_{k}^{(n)}$'s are independent and $\mathbb{E}_{\mathcal{Q}}\left[\left\Vert\mathcal{Q}(\mathbf{g}_{k}^{(n)})-\mathbf{g}_{k}^{(n)}\right\Vert\right]^{2} \leq \frac{\sqrt{d}}{q}\left\Vert\mathbf{g}_{k}^{(n)}\right\Vert^{2}$ holds, it follows that 
\begin{equation*}
	\mathbb{E}_{\mathcal{Q}}[\left\Vert \mathbf{a}_{2}\right\Vert^{2}] \leq \frac{\sqrt{d}\eta_{n}^{2}}{qK^{2}}\sum_{k=1}^{K}\left\Vert\mathbf{g}_{k}^{(n)}\right\Vert^{2}.
\end{equation*}

\subsection{\textbf{Bound of (\ref{eq:w_t_plus_1})}}

With these bounds at hand, taking expectation of (\ref{eq:w_t_plus_1}) over the stochastic quantizer $\mathcal{Q}$ and stochastic gradient at round $n$, we have that 
\begin{align}
	\nonumber &\mathbb{E}\left[\left\Vert \mathbf{w}^{(n+1)} - \mathbf{w}_{\ast} \right\Vert^{2}\right] \\
	&\nonumber \leq \left(1 - \mu\eta_{n}\right)\left\Vert \mathbf{w}^{(n)} - \mathbf{w}_{\ast} \right\Vert^2\\
	& \nonumber + 2\frac{\eta_{n}}{K}\sum_{k=1}^{K}\left(F_{k}(\mathbf{w}_{\ast}) - F_{k}(\mathbf{w}^{(n)})\right)\\
	&
	\label{eq:w_t_plus_1_renew}+\frac{\alpha\eta_{n}^{2}}{K}\sum_{k=1}^{K}\mathbb{E}\Vert\mathbf{g}_{k}^{(n)}\Vert^{2}.
\end{align}
Recall that $\alpha=\frac{\sqrt{d}}{qK}+1$. From Assumption \ref{asp:gradient}, we have that 
\begin{equation}\label{eq:gradient_norm_square}
	\mathbb{E}\Vert\mathbf{g}_{k}^{(n)}\Vert^{2} \leq \delta_{k}^{2} + \Vert \nabla F_{k}(\mathbf{w}^{(n)}) \Vert^{2}.
\end{equation}
After substituting (\ref{eq:gradient_norm_square}) into (\ref{eq:w_t_plus_1_renew}), it yields
\begin{align}
	\nonumber &\mathbb{E}\left[\left\Vert \mathbf{w}^{(n+1)} - \mathbf{w}_{\ast} \right\Vert^{2}\right] \\
	& \nonumber \leq \left(1 - \mu\eta_{n}\right)\left\Vert \mathbf{w}^{(n)} - \mathbf{w}_{\ast} \right\Vert^2 \\
	& \nonumber + 2\frac{\eta_{n}}{K}\sum_{k=1}^{K}\left(F_{k}(\mathbf{w}_{\ast}) - F_{k}(\mathbf{w}^{(n)})\right)\\
	&
	\nonumber+\frac{\alpha\eta_{n}^{2}}{K}\sum_{k=1}^{K}\left(\delta_{k}^{2} + \Vert \nabla F_{k}(\mathbf{w}^{(n)}) \Vert^{2}\right).
\end{align}
The $L$-smoothness of $F_{k}(\mathbf{w})$ gives that
\begin{equation*}
	\Vert\nabla F_{k}(\mathbf{w}^{(n)})\Vert^{2} \leq 2L\left(F_{k}(\mathbf{w}^{(n)})-F_{k}^{\ast}\right),
\end{equation*}
It follows that
\begin{align}
	&\nonumber \mathbb{E}\left[\left\Vert \mathbf{w}^{(n+1)} - \mathbf{w}_{\ast} \right\Vert^{2}\right] \\
	&\nonumber \leq \left(1 - \mu\eta_{n}\right)\left\Vert \mathbf{w}^{(n)} - \mathbf{w}_{\ast} \right\Vert^2 + \frac{\alpha\eta_{n}^{2}}{K}\sum_{k=1}^{K}\delta_{k}^{2}\\&
	\nonumber + \underbrace{2\frac{\eta_{n}}{K}\sum_{k=1}^{K}\left(F_{k}(\mathbf{w}_{\ast}) - F_{k}(\mathbf{w}^{(n)})\right)}_{C_{1}}\\
	&+\nonumber \underbrace{\frac{2L\eta_{n}^{2}\alpha}{K}\sum_{k=1}^{K}\left(F_{k}(\mathbf{w}^{(n)})-F_{k}^{\ast}\right)}_{C_{2}}.
\end{align}
After rearranging $C_{1}+C_{2}$, it becomes
\begin{align*}
	C_{1} + C_{2} &= 2\eta_{n}\left(\alpha L\eta_{n}-1\right)\left(F(\mathbf{w}^{(n)})-F(\mathbf{w}_{\ast})\right) \\
	&+ 2\alpha L\eta_{n}^{2}F_{\delta},
\end{align*}
where $F_{\delta} \coloneqq F(\mathbf{w}_{\ast}) - \frac{1}{K}\sum_{k=1}^{K}F_{k}^{\ast}$.
It can be verified that $\eta_{n} \leq \frac{1}{\alpha L}$, and from $F(\mathbf{w}^{(n)}) \ge F(\mathbf{w}_{\ast})$, we have
\begin{equation*}
	C_1 \leq 2\alpha L\eta_{n}^{2}F_{\delta}.
\end{equation*}
Taking total expectation of (\ref{eq:w_t_plus_1}), it yields that
\begin{align*}
	\mathbb{E}\left[\left\Vert \mathbf{w}^{(n+1)} - \mathbf{w}_{\ast} \right\Vert^{2}\right] &\leq \left(1 - \mu\eta_{n}\right)\mathbb{E}\left[\left\Vert \mathbf{w}^{(n)} - \mathbf{w}_{\ast} \right\Vert^2\right] \\
	&+ \alpha\eta_{n}^{2}(2 LF_{\delta} +\frac{1}{K}\sum_{k=1}^{K}\delta_{k}^{2}),
\end{align*}
which completes the proof.

\section{Proof of Lemma \ref{prop}}\label{apdx:prop_1}

If $T_{k}^{\text{comp}}+T_{k}^{\text{comm}} < T_{d}$, $\forall k\in[K]$, $T_{d}$ can be reduced until some $k \in [K]$ satisfy that $T_{k}^{\text{comp}}+T_{k}^{\text{comm}} = T_{d}$. Denote $\mathcal{K}=\{k\in[K] \vert T_{k}^{\text{comp}}+T_{k}^{\text{comm}} < T_{d}\}$ and $\bar{\mathcal{K}} = \{k\in[K] \vert T_{k}^{\text{comp}}+T_{k}^{\text{comm}} = T_{d}\}$. Obviously, $\mathcal{K} + \bar{\mathcal{K}} = [K]$. Since $T_{k}^{\text{comm}}$ is an decreasing function of $b_{k}$, we can enforce $T_{k}^{\text{comp}}+T_{k}^{\text{comm}} = T_{d}$ by decreasing $b_{k}$ for all $k\in \mathcal{K}$. Then, $\mathcal{K}=\emptyset$ and $\bar{\mathcal{K}}=[K]$. In this case, if $\sum_{k=1}^{K}b_{k} < B$, we can properly increase each $b_{k}$, without violating $T_{k}^{\text{comp}}+T_{k}^{\text{comm}} = T_{d}$, $k\in[K]$, until $\sum_{k=1}^{K}b_{k} = B$, and $T_{d}$ will decrease as well.
		
\bibliographystyle{IEEEtran}
\bibliography{bibsample}

\end{document}